\documentclass{ptephy_v1}

\usepackage[utf8]{inputenc}

\usepackage{amsmath} 
\usepackage{amsthm} 
\usepackage{graphics} 
\usepackage{algorithmic}
\usepackage{subfig} 
\usepackage{url} 

\usepackage{bm}
\usepackage{mathrsfs}
\usepackage{rsfso}
\usepackage{array}
\usepackage{tabularx}

\usepackage{hhline}
\usepackage{ascmac}
\usepackage{graphicx}
\usepackage{multirow}

\usepackage{wrapfig}
\usepackage{comment}
\usepackage{color}

\begin{document}

\title{First limits from a 
3d-vector directional dark matter search   
with the NEWAGE-0.3b' detector}

\author[1]{Ryota~Yakabe}
\author[1]{Kiseki~Nakamura}
\author[1]{Tomonori~Ikeda}
\author[1]{Hiroshi~Ito}
\author[1]{Yushiro~Yamaguchi}
\author[1]{Ryosuke~Taishaku}
\author[1]{Miki~Nakazawa}
\author[1]{Hirohisa~Ishiura}
\author[1]{Takuma~Nakamura}
\author[1]{Takuya~Shimada}
\author[2]{Toru~Tanimori}
\author[2]{Hidetoshi~Kubo}
\author[2]{Atsushi~Takada}
\author[3,4]{Hiroyuki~Sekiya}
\author[3,4]{Atsushi~Takeda}
\author[1]{Kentaro~Miuchi}

\affil[1]{Department of Physics, Graduate School of Science, Kobe University, 1-1 Rokkodai-cho, Nada-ku, Kobe, Hyogo, 657-8501, Japan \email{miuchi@phys.sci.kobe-u.ac.jp}}
\affil[2]{Division of Physics and Astronomy, Graduate School of Science, Kyoto University, Kitashirakawaoiwake-cho, Sakyo-ku, Kyoto, Kyoto, 606-8502, Japan}
\affil[3]{Kamioka Observatory, Institute for Cosmic Ray Research, the University of Tokyo, Higashi-Mozumi, Kamioka, Hida, Gifu, 506-1205, Japan}
\affil[4]{Kavli Institute for the Physics and Mathematics of the Universe (WPI), the University of Tokyo, 5-1-5 Kashiwanoha, Kashiwa, Chiba, 277-8582, Japan}


\begin{abstract}
The first directional dark matter search with three-dimensional tracking with head-tail sensitivity (3d-vector tracking analysis) was performed with a gaseous three-dimensional tarcking detector, or the NEWAGE-0.3b' detector.
The search was carried out from July 2013 to August 2017 (Run14 to Run18) at the Kamioka underground laboratory.
The total livetime is 434.85~days corresponding to an exposure of 4.51~kg$\cdot$days.
A 90~\% confidence level upper limit on spin-dependent WIMP-proton cross section of $4.3\times10^{2}$~pb for WIMPs with the mass of 150~GeV/$c^2$ is obtained.
\end{abstract}

\subjectindex{Dark matter, $\mu$-TPC}


\maketitle
\section{Introduction}
\label{sec:intro}
A considerable number of cosmological observations show a strong evidence that an unknown particle, so called dark matter, constitutes about 27\% of the universe\cite{1}.
Weakly interacting massive particles (WIMPs) are considered to be one of the best dark matter candidates and direct detection experiments have sought for the evidence of the elastic scattering between a WIMP and a nucleus. 
With a natural assumption that the dark matter is gravitationally trapped  in galaxies, the solar system should receive a dark matter wind due to the rotation around the center of Milky Way. 
Detecting the direction of nuclear recoil tracks has been considered to be a reliable detection method for positive WIMP signatures\cite{2,3}.
Furthermore, it is said that this method works as a strong tool in the search for WIMPs even below the so-called neutrino floor at which the WIMP search sensitivity starts to be limited by neutrino-nucleus coherent scatterings\cite{4}.
Gaseous time projection chambers (TPCs) can detect the directions of nuclear recoil tracks. Several types of TPCs with both charge and optical readout have been developed and dark matter searches are carried out\cite{5,6,7}.

NEWAGE is a directional direct dark matter search experiment using 
a three-dimensional gaseous tracking detector, or a micro time projection chamber ($\mu$-TPC).
Here we refer to a time projection chamber read by micro-patterned gaseous detectors as $\mu$-TPC\cite{8}.
The latest limits on WIMP-proton cross-section by NEWAGE are the results from a measurement of about 30~days \cite{6}.
We continued the measurement keeping the same detector condition and increased the statistics by a fact of more than ten.
We also update the analysis so that the senses of nuclear recoil tracks (head-tails) can statistically be known with the asymmetric charge deposition of the 
nuclear recoil tracks \cite{9}.

In this paper, 
the first directional dark matter search with three-dimensional tracking with head-tail sensitivity (3d-vector tracking analysis) performed with the NEWAGE-0.3b' is described.
The detector system and its performance including the event selection are described in Sec.~2.
In Sec.~3, the measurement properties and results of the search are described.
Future prospects are discussed in Sec.~4 and the paper is concluded in Sec.~5.



\section{Detector}
\label{sec:newage}

\subsection{NEWAGE-0.3b' overview}
\label{sec:03b}
\begin{figure}[htbp]
    \begin{center}
       \includegraphics[height=70mm,width=120mm]{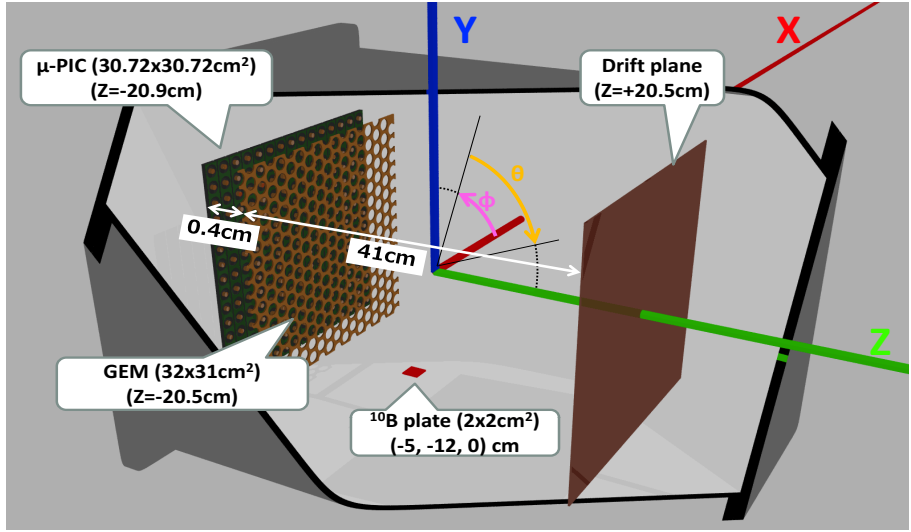}
    \end{center}
    \caption{Schematic view of the $\rm \mu$-TPC. The $\rm \mu$-TPC consists of a $\rm \mu$-PIC, a GEM, and a drift cage. The detection volume is  $30.7 \times 30.7 \times 41.0~\rm cm^3$.}
    \label{fig:TPC}
\end{figure}

One of the NEWAGE detectors, NEWAGE-0.3b', was used for this directional WIMP search\cite{6}. 
NEWAGE-0.3b' consists of a micro time projection chamber ($\mu$-TPC), 
a gas circulation system, and a readout electronics system.
Schematics of the $\mu$-TPC and its internal structure are shown in Fig.~\ref{fig:TPC}.
The $\mu$-TPC consists of 
a micro pixel chamber ($\mu$-PIC) \cite{10}, a gas electron multiplier (GEM) \cite{11}, and a drift cage. The detection volume is 30.7 $\times$ 30.7 $\times$ 41.0~$\rm cm^3$. The X and Y axes are defined to be parallel to the $\mu$-PIC readout strips and the Z-axis is defined to be parallel to the drift direction.
The origin of the axis is set at the center of the detection volume.
The detection volume was filled with $\rm CF_4$ gas at 76~torr. 
$\rm CF_4$ gas is selected as the target gas because the gas diffusion is small and fluorine has a relatively large spin-dependent (SD) cross section for WIMPs\cite{12}.

A $\mu$-PIC (Dai Nippon Printing Co., Ltd.) is one of the micro-patterned gaseous detectors and is manufactured using printed circuit board (PCB) technology \cite{10}.
The PCB technology can produce a large-sized detector at a reasonable cost, which is one of the most important requirements for WIMP search detectors
The effective area of the $\mu$-PIC for the NEWAGE-0.3b' is 30.7 $\times$ 30.7~$\rm cm^2$ read by two-dimensional strips with a pitch of 400 $\mu$m in both the X and Y directions.
Because of the structure of the electrodes, these two-dimensional strips are referred to as anode ($x$) and cathode ($y$) strips, respectively.
Hereafter, the strip IDs are represented by $x$ and $y$, whereas the positions in real geometry are expressed by $X$ and $Y$ in units of centimeters.
Positive bias is applied to the anode electrodes which are in a
shape of pixels with an outer diameter of 70~$\rm \mu m$. Gas amplification takes place around the anode electrodes.
Ions drift towards the cathode electrodes which have a circular shape of inner diameter of 260~$\mu$m. 
Same amount of positive and negative charges
are read through the anode and cathode strips, respectively.
A GEM, manufactured by SciEnergy., Ltd. is placed in between the $\rm \mu$-PIC and the detection volume. The distance between the GEM and the $\rm \mu$-PIC is 0.4~cm. 
The GEM performs as a first-stage electron-amplifier to ensure gas gain while reducing the risk of the discharges at the $\rm \mu$-PIC.
The effective area of the GEM (31 $\times$ 32~$\rm cm^2$) covers the entire detection area of the $\rm \mu$-PIC.  
The GEM is made of a 100~$\rm \mu$m-thick liquid crystal polymer and the hole size and pitch are 70~$\rm \mu$m and 140~$\rm \mu$m, respectively. 
The drift length is 41.0~cm.
The electric field is formed by a drift plane and 1 cm-spaced wires on side walls made of polyetheretherketone. 
A glass plate with a thin layer of $^{10}\mathrm{B}$ is installed at a position of $(X, Y, Z)=(-5.0, -12.0, 0.0)$ for energy calibration. The size of the $^{10}\mathrm{B}$ layer is 2.0 $\times$ 2.0~$\rm cm^2$ with a thickness of 0.6~$\mu$m.
By irradiating the $^{10}\mathrm{B}$ plate with thermalized neutrons from a $^{252}\mathrm{Cf}$ source  surrounded by polyethylene blocks, alpha-rays are generated by the following reactions.

\begin{eqnarray}
\label{eq:10B1}
^{10}\mathrm{B} + n & \rightarrow & ^{4}\mathrm{He} + ^{7}\mathrm{Li} + 2.79 \ {\rm MeV} (6\%)\\
\label{eq:10B2}
^{10}\mathrm{B} + n & \rightarrow & ^{4}\mathrm{He} + ^{7}\mathrm{Li} + 2.31 \ {\rm MeV} + \gamma\ (0.48 \ {\rm MeV}) (94\%)
\end{eqnarray}
The kinetic energy of the alpha-rays in reaction (\ref{eq:10B2}), 1.5MeV, is effectively observed.
The $\mu$-TPC is placed in a stainless-steel vacuum vessel.

A gas circulation system with cooled charcoal is used to reduce 
radon gas which is a major source of background for WIMP searches, and to maintain the gas quality during long-term measurements.
The gas in the vessel passes through the filter where 100~g of charcoal (TSURUMICOAL 2GS) absorbes the radon and other impurities.
The gas is circulated at a rate of 500 $\sim$ 1000 mL/min using a dry pump (XDS5 Scroll Pump, EDWARDS).
Stable cooling at 230 K $\pm 2~\%$ is realized by controlling a heater, whereas the cooler (CT-910 Cool Man Trap (SIBATA)) is always operated at its maximum cooling power.
The detector stability will be discussed in Sec.~\ref{detector_sta}.

\begin{figure}[htbp]
    \begin{center}
       \includegraphics[width=150mm]{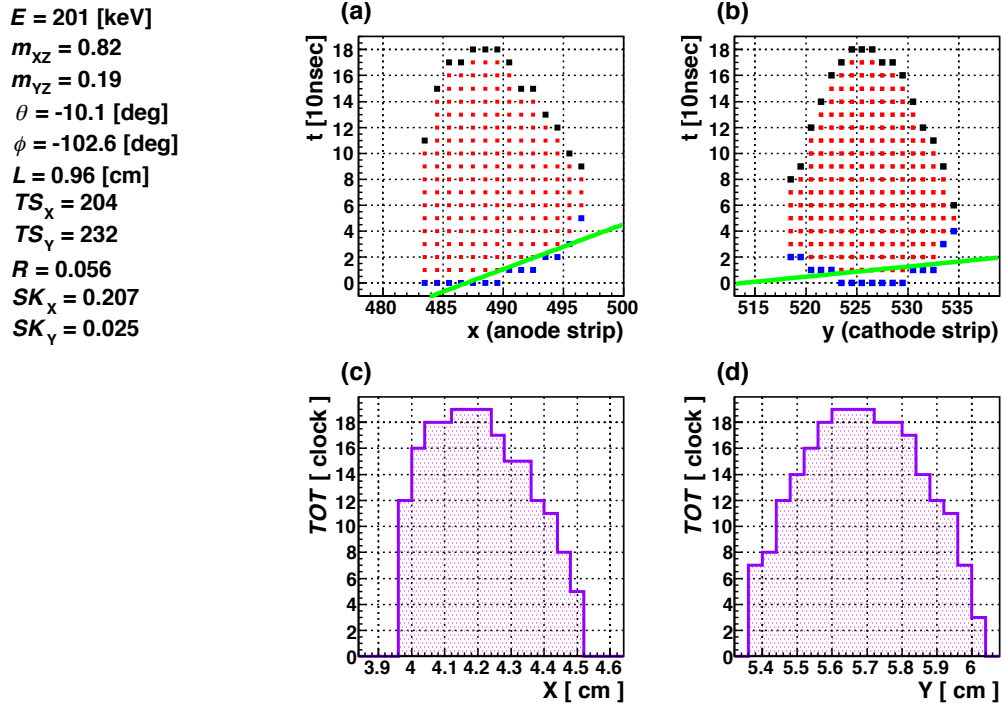}
    \end{center}
    \caption{A typical event display for a measurement with a $^{252}\mathrm{Cf}$ neutron source. The top-left (a) and top-right (b) panels show raw track data on the $x-t$ and $y-t$ panels, respectively.
      Blue and black markers show the rising and falling edges, respectively. Red markers show the durations of the signals. Green lines show the fitted track. The bottom-left (c) and bottom-right (d) panels show time-over-threshold ($TOT$) distributions along the $X$ and $Y$ axes, respectively. Details are described in Sec.~\ref{sec:03b} and Sec~\ref{sec:param}.
      \label{fig:EvtDis}}
\end{figure}

A data acquisition (DAQ) system dedicated to the $\mu$-PIC readout was used for the measurement\cite{6}.
The DAQ system recorded two types of data, ``charges" by a flash-ADC (FADC) and ``tracks" by a memory board.
Analog signals from 768~cathode electrodes are grouped down to 4~channels and their waveforms are recorded with a 100~MHz FADC as the charge data.
The rising and falling edge timings ($t$) of each strip are recorded as the track data.
TPC analog signals from 768~anode strips are grouped down to 16 channels and a hit at any one of them is used as a trigger. 
In this DAQ system, the drift length or the absolute $Z$ position is not measured since the trigger is issued at the arrival timing of the electrons on the $\mu$-PIC, not at the actual time of the event.
If there is any way to know the actual event time, like an external trigger, the 
absolute $Z$ position can be known from the time difference between the actual event time and the electrons’ arrival time on the detection plane.
However, there is no established way to know the actual event time for this type of self-triggering TPC, thus the absolute $Z$ position is not known. 
A typical nuclear recoil track sample taken with this DAQ system 
is shown in Fig.~\ref{fig:EvtDis}.
Several parameters are defined to characterize the track property of each event.

\subsection{Event parameters}
\label{sec:param}
Waveforms recorded by the FADC are used to estimate the charge of each event.
The charge of each event is calculated by integrating the waveform over the time duration when the voltage exceeded the threshold.
The charge is then converted into energy ($E$) with a calibration factor. 
Energy calibration is carried out with $\alpha$-rays using the $\rm {}^{10}B(n,\alpha){}^{7}Li$ reaction by irradiating the $\rm{}^{10}B$ plate with thermal neutrons (Eqs. (\ref{eq:10B1}) and (\ref{eq:10B2}))[6].
The ionization efficiency is corrected with SRIMs simulations\cite{13}.
The energy scale calibrated by $\alpha$-rays is used for further discussion.

Event parameters on the track information will be explained using an event display shown in  Fig.~\ref{fig:EvtDis}.
The upper panels ((a),(b)) show the recorded raw data on the anode ($x$) and cathode ($y$) strips, respectively.
The rising edge (shown with blue marker) and the falling edge (shown with black marker) of each strip are recorded and the time-over-threshold ($TOT$) of each strip is determined as the difference between the clocks of these edges. 
The rising edge represents the arrival time of drifting electrons on each strip.
The relative times of rising edges to the trigger timing thus represent the relative $Z$ positions within a track.
The rising edges in Fig.~\ref{fig:EvtDis} are fitted with straight lines in the $x-t$ and $y-t$ planes. The best fit ones are shown with green lines.
The $TOT$ of each strip corresponds to the energy deposition therein.
The $TOT$s of strip IDs $x$ and $y$ are defined as $TOT(x)$ and $TOT(y)$, respectively.
The $TOT$-sum of the $x$ strips ($TS_{\rm x}$) and $y$ strips ($TS_{\rm y}$) are defined as Eqs.~(\ref{eq:TOT_X}) and~(\ref{eq:TOT_Y}).

\begin{eqnarray}
TS_{\rm x} &\equiv& \sum_{x=x_{\rm min}}^{x_{\rm max}} TOT(x) \label{eq:TOT_X}\\
TS_{\rm y} &\equiv& \sum_{y=y_{\rm min}}^{y_{\rm max}} TOT(y) \label{eq:TOT_Y},
 \end{eqnarray}
 
where $\rm min$ and $\rm  max$ represent the minimum and maximum IDs of the hit strips on the corresponding coordinate. 

Lower panels ((c),(d)) of Fig.~\ref{fig:EvtDis} show the $TOT$s with rising edges shifted to zero for a better view of the $TOT$ distributions along the axes.
To parameterize the asymmetry of the $TOT$ distribution, or the energy deposition, along the $x$ and $y$ coordinates,
a parameter named skewness is defined.
Here the energy deposition distribution has some information about the track sense along each axis.
The tracks 
in the energy range of interest for this study (below 400~keV) 
are known to have larger energy depositions at their beginnings than at their ends\cite{14}. 
In the event shown in Fig.~\ref{fig:EvtDis}, the track ran from left to right on the $X$ axis, whereas no clear difference is not seen along the $Y$ axis. 
Skewnesses ($SK_{\rm x}$ and $SK_{\rm y}$) for the $x$ and $y$ strips are defined by Eqs.~(\ref{eq:skewness1}) - ~(\ref{eq:skewness6}).

\begin{eqnarray}
 SK_{\rm x} &=& \frac{S_{3x}}{S^{3/2}_{2x}},\label{eq:skewness1}\\
 S_{nx} &\equiv& {\sum_{x=x_{\rm min}}^{x_{\rm max}}} \frac{(x - <x>)^n \cdot TOT(x)}{TS_{\rm x}},\label{eq:skewness2}\\ 
 <x>& =& {\sum_{x=x_{\rm min}}^{x_{\rm max}}} \frac{x \cdot TOT(x)}{TS_{\rm x}},\label{eq:skewness3}\\
 SK_{\rm y} &=& \frac{S_{3y}}{S^{3/2}_{2y}},\label{eq:skewness4}\\
 S_{ny} &\equiv& {\sum_{y=y_{\rm min}}^{y_{\rm max}}} \frac{(y - <y>)^n \cdot TOT(y)}{TS_{\rm y}},\label{eq:skewness5}\\ 
 <y>& =& {\sum_{y=y_{\rm min}}^{y_{\rm max}}} \frac{y \cdot TOT(y)}{TS_{\rm y}}.\label{eq:skewness6} 
 \end{eqnarray}

Until now, the parameters in the $x-t$ and $y-t$ planes has been used for the discussion, where $t$ is defined in units of time.
With drift velocity information measured as one of the calibration parameters, $t$ can be converted into $Z$ which has units of length.
Hereafter, the parameters are defined in the $X-Z$ and $Y-Z$ planes with conversions into $X$ and $Y$ units. 
To characterize the track shapes, the best-fit lines on the rising edges in the $x-t$ and $y-t$ planes shown with green lines in Fig.~\ref{fig:EvtDis} are converted to the $X-Z$ and $Y-Z$ planes.
Here the slopes of the best-fit lines in the $X-Z$ and $Y-Z$ planes are parameterized as $m_{\rm X}$ and $m_{\rm Y}$, respectively.
Converted lines to the $X-Z$ and $Y-Z$ planes are expressed with $Z = m_{\rm X}\cdot X + n_{\rm X}$ and $Z = m_{\rm Y} \cdot Y +  n_{\rm Y}$, respectively. Here $m_{\rm X}$ and $m_{\rm X}$ are slopes and $n_{\rm X}$ and $n_{\rm Y}$ are Z-axis sections.

The length of the best-fit line between the $\rm min$ and $\rm max$ strip ID is calculated on each plane.
$\Delta  X(\Delta Y)$ is the projection on $X (Y)$ axis of the track line of the $X-Z$ ($Y-Z$) plane.
$\Delta Z$s are  accordingly defined as $m_{\rm X}\cdot \Delta X$  and $m_{\rm Y}\cdot \Delta Y$, in the $X-Z$ plane and in $Y-Z$ plane, respectively.
The smaller one of the $\Delta Z$s is then used as $\Delta Z$. 
The length in the 3D space ($L$) is calculated by $L=\sqrt{\Delta X^2+\Delta Y^2+\Delta Z^2}$.

The roundness parameter ($R$), which represents the shape of the tracks, is defined by Eq.~(\ref{eq:roundnessX})~-~(\ref{eq:roundness}).

\begin{eqnarray}
      {R_ {\rm X}}  & = & \sum^{X_{\rm max}}_{X=X_{\rm min}} \frac{ (Z_{\rm rise}(X) - m_{\rm X}X - n_{\rm X})^2}{X_{\rm max}-X_{\rm min}}, \label{eq:roundnessX}\\
      {R _{\rm Y}} &= & \sum^{Y_{\rm max}}_{Y=Y_{\rm min}} \frac{ (Z_{\rm rise}(Y) - m_{\rm Y}Y - n_{\rm Y})^2}{Y_{\rm max}-Y_{\rm min}} , \\
 {R} & = & {\rm min({\it R}_{\rm X}, {\it R}_{\rm Y})}, \label{eq:roundness}
\end{eqnarray}
where $Z_{\rm rise}(X)$ and $Z_{\rm rise}(Y)$ are the rising edges at $X$ and $Y$, respectively.
The $SK$ parameters are introduced to express the asymmetry of nuclear recoil tracks for the head-tail recognition.
The $R$ parameter is introduced to reject gamma-BG\cite{6}.

Because $m$ and $SK$ are found to have correlations due to the time-walk effect of the rising-edge timing, these parameters are corrected so that $m$ values would not show $SK$ dependence in the $m-SK$ planes.
The corrections are the rotations of $m_{\rm X}$ and $SK_{\rm X}$ in $m_{\rm X}-SK_{\rm X}$ plane and $m_{\rm Y}$ and $SK_{\rm Y}$ in $m_{\rm Y}-SK_{\rm Y}$ plane.
Corrected parameters ($cm_{\rm X}, cm_{\rm Y}, cSK_{\rm X}$, and $cSK_{\rm Y}$) are used for further discussions.
The azimuth($\phi$) and elevation ($\theta$) of the tracks are also calculated from the corrected slope parameters ($cm_{\rm X}$ and $cm_{\rm Y}$).

\subsection{Event Selection and Detector Performance}
\label{sec:evtsel}
Several event selections are applied to reject various types of backgrounds.
A fiducial area of 28.0 $\times$ 24.0~$\rm cm^2$ is defined and outer area is used as a veto region to discriminate background events.
Main background events rejected by this veto are protons originating from the wall of the TPC field cage.
This area is not symmetric because of the $^{10}\mathrm{B}$ plate.
We need an additional veto region around the $^{10}\mathrm{B}$ plate.
This cut (fiducial cut) requires that the whole track is in this fiducial area.

The electron tracks mainly due to external $\gamma$-rays are rejected by cuts using the parameters $L$ and $TS$.
We refer to them ``L cut" and ``TS cut", respectively.
We refer to any cut using parameter ``$A$” as a ``A cut” in the following discussion.
This rejection is realized based on the fact that the energy deposition per unit distance of an electron is much smaller than that of a nucleus. Therefore, the $L$ and $TS$ of the electrons are expected to be longer and smaller than those of the nucleus, respectively. 
In this work, an energy dependence is taken into account to the TS cut to improve the rejection power.

Alpha-ray events from the $\mu$-PIC, which are a major source of backgrounds, are eliminated by the R cut.
These four cuts are also used in our previous work\cite{6}. 
Two additional cuts, the $\theta$ cut and the SK cut, are introduced for this work.
The $\theta$ cut rejected $\alpha$-ray background going through GEM holes from the $\mu$-PIC.
These tracks have large $\theta$ so a cut of large $\theta$ would discriminate these events.
The SK cut is introduced to enhance the head-tail discrimination power by rejecting events with small absolute values of $SK$s.  
The cut criteria are listed below.

\begin{itemize}
 \setlength{\itemsep}{0.3 cm}
  \item Fiducial cut :  ($X$/cm) $\leq$ $-14$ or 14 $\leq$ ($X$/cm) or ($Y$/cm) $\leq$  $-10$ or 14 $\leq$ ($Y$/cm) 
 \item L cut : ($L$/cm) $>$ 0.6 + 0.004 $\times$ ($E$/keV) 
 \item TS cut : $TS_{\rm X} <$ 50 + 0.5 $\times (E$/keV) or $TS_{\rm Y} <$ 50 + 0.5 $\times (E$/keV)
 
 \item R cut : $R/{\rm cm^2} < 0.04 $
 \item $\rm \theta$ cut : $\mid$ sin$\theta \mid$ $<$ 0.5 
 \item SK cut : $\mid$ $cSK$ $\mid$ $<$ 0.1 
\end{itemize}

\begin{figure}[htbp]
  \begin{center}
    \begin{tabular}{c}

      \begin{minipage}{0.5\hsize}
        \begin{center}
          \includegraphics[width=7cm]{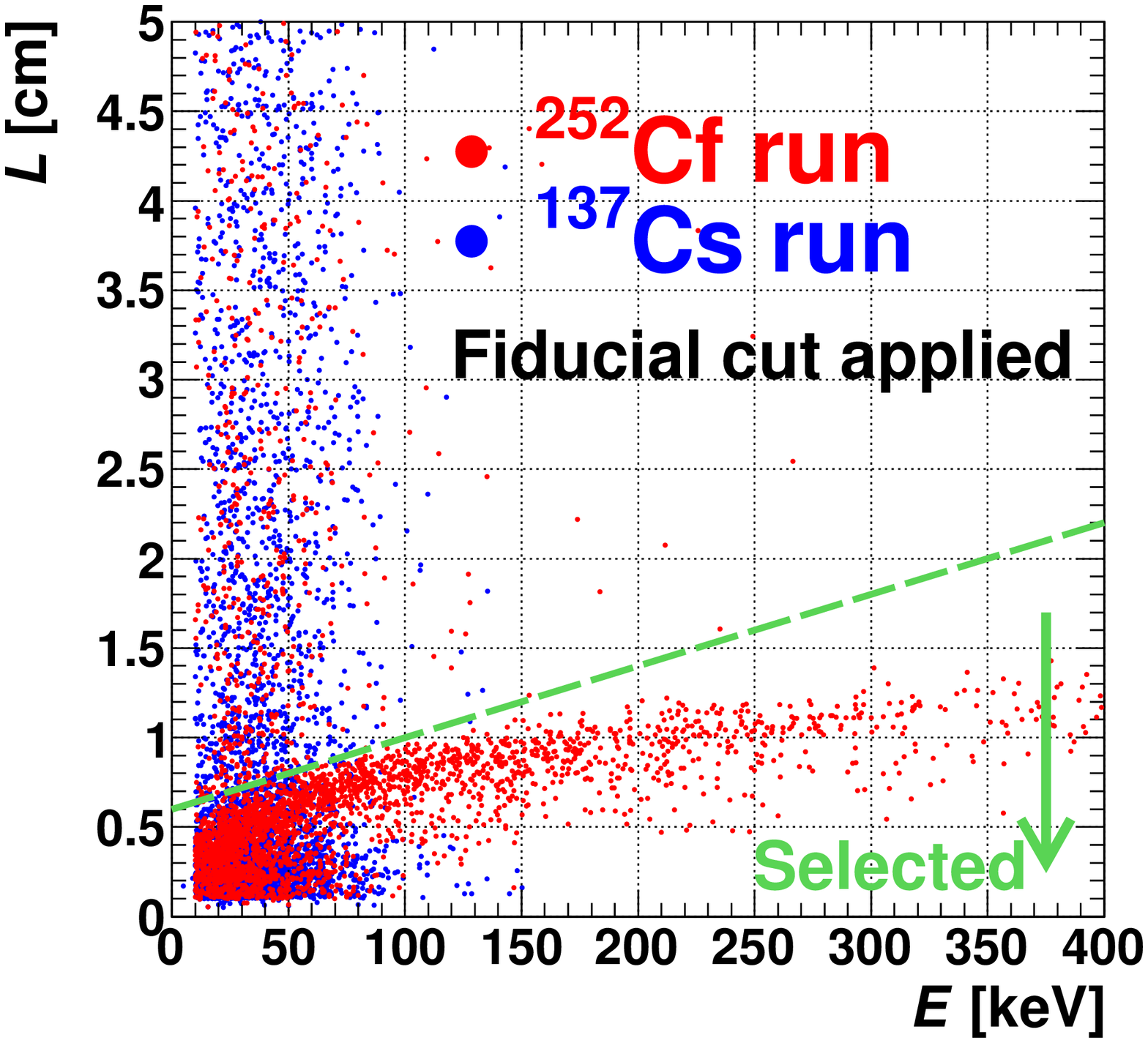}
          \hspace{1.6cm}
        \end{center}
      \end{minipage}

      \begin{minipage}{0.5\hsize}
        \begin{center}
          \includegraphics[width=7cm]{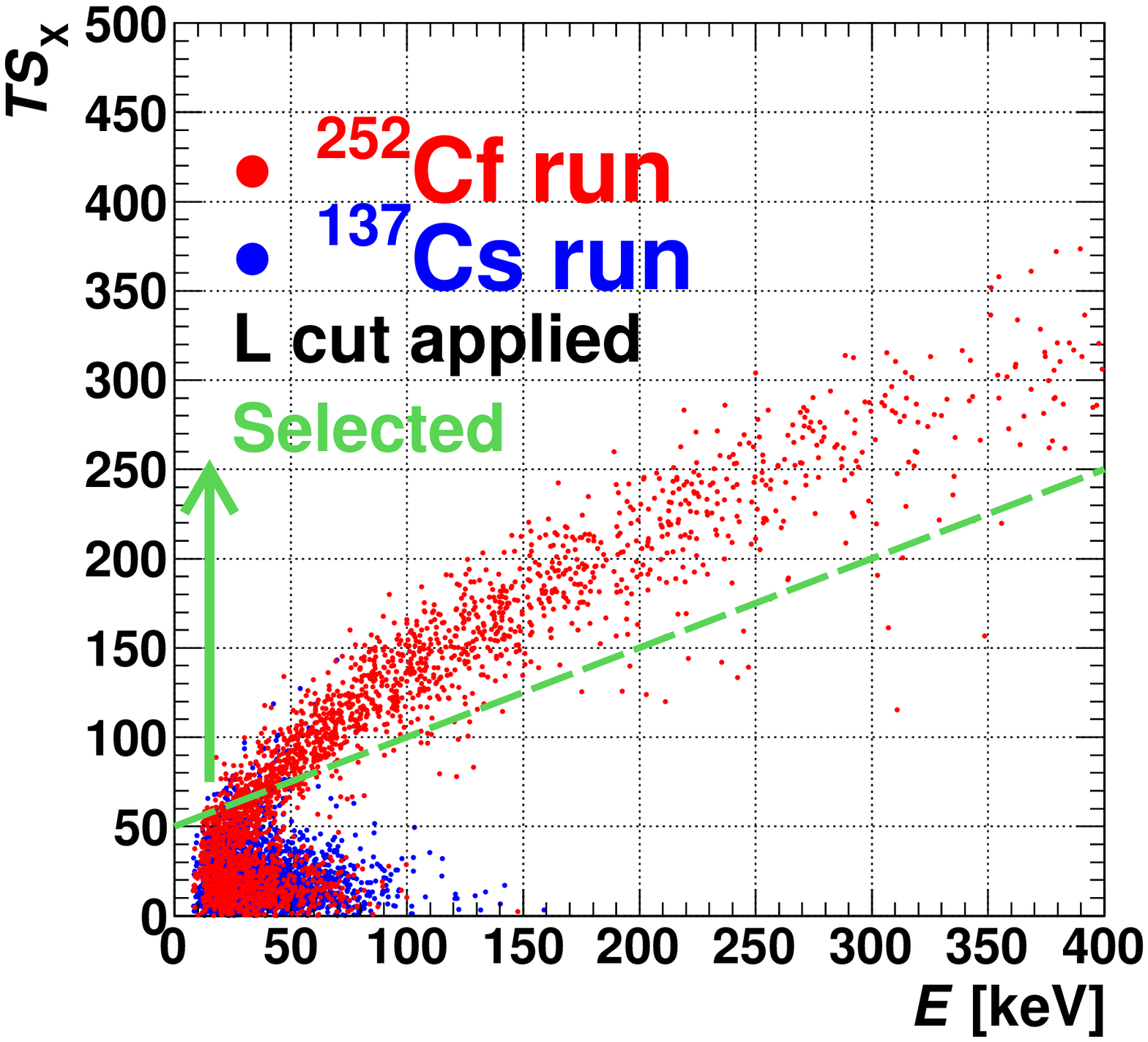}
          \hspace{1.6cm}
        \end{center}
      \end{minipage}\\
    
    \begin{minipage}{0.5\hsize}
      \begin{center}
        \includegraphics[width=7cm]{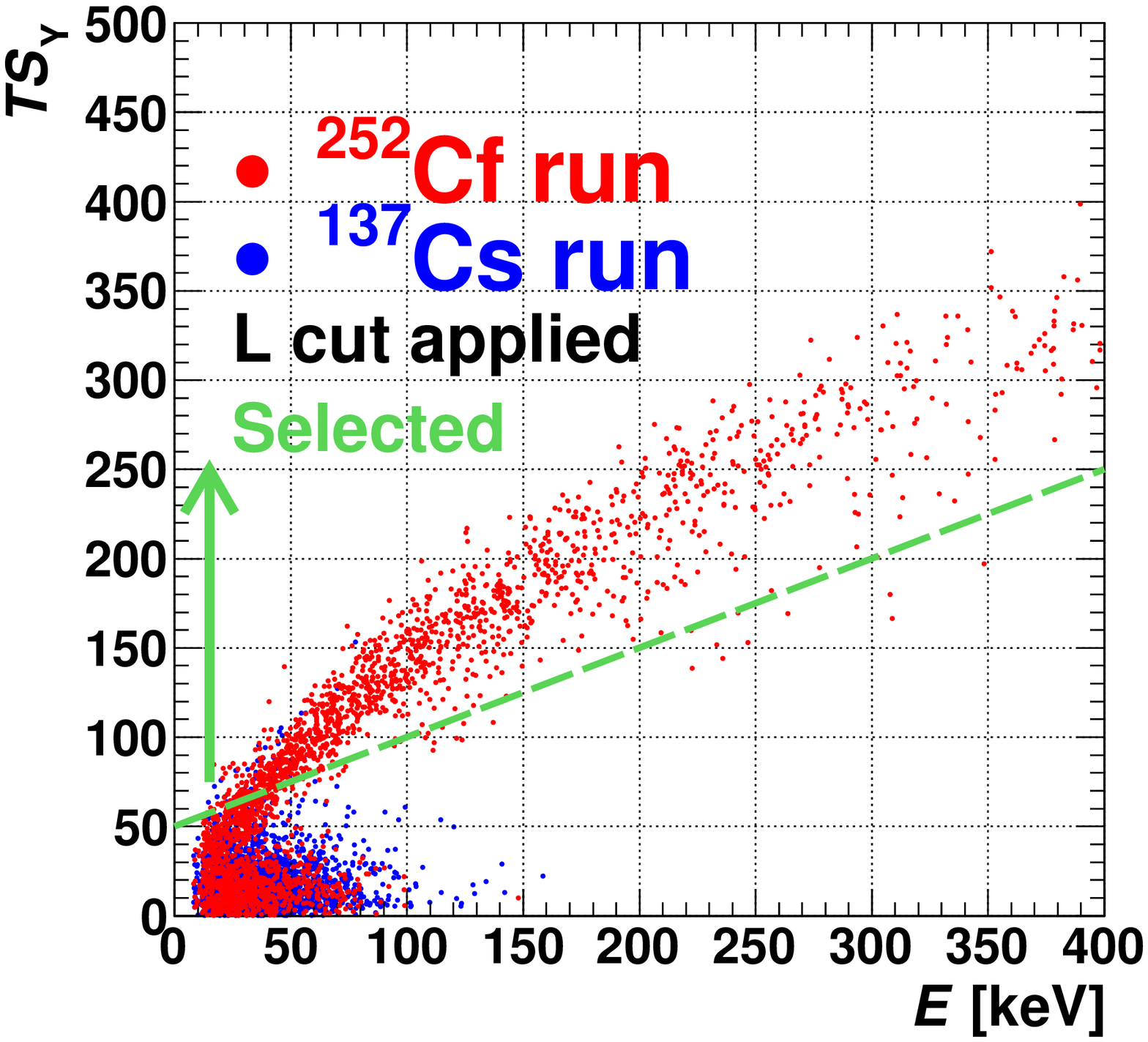}
        \hspace{1.6cm}
      \end{center}
    \end{minipage}
      
    \begin{minipage}{0.5\hsize}
      \begin{center}
        \includegraphics[width=7cm]{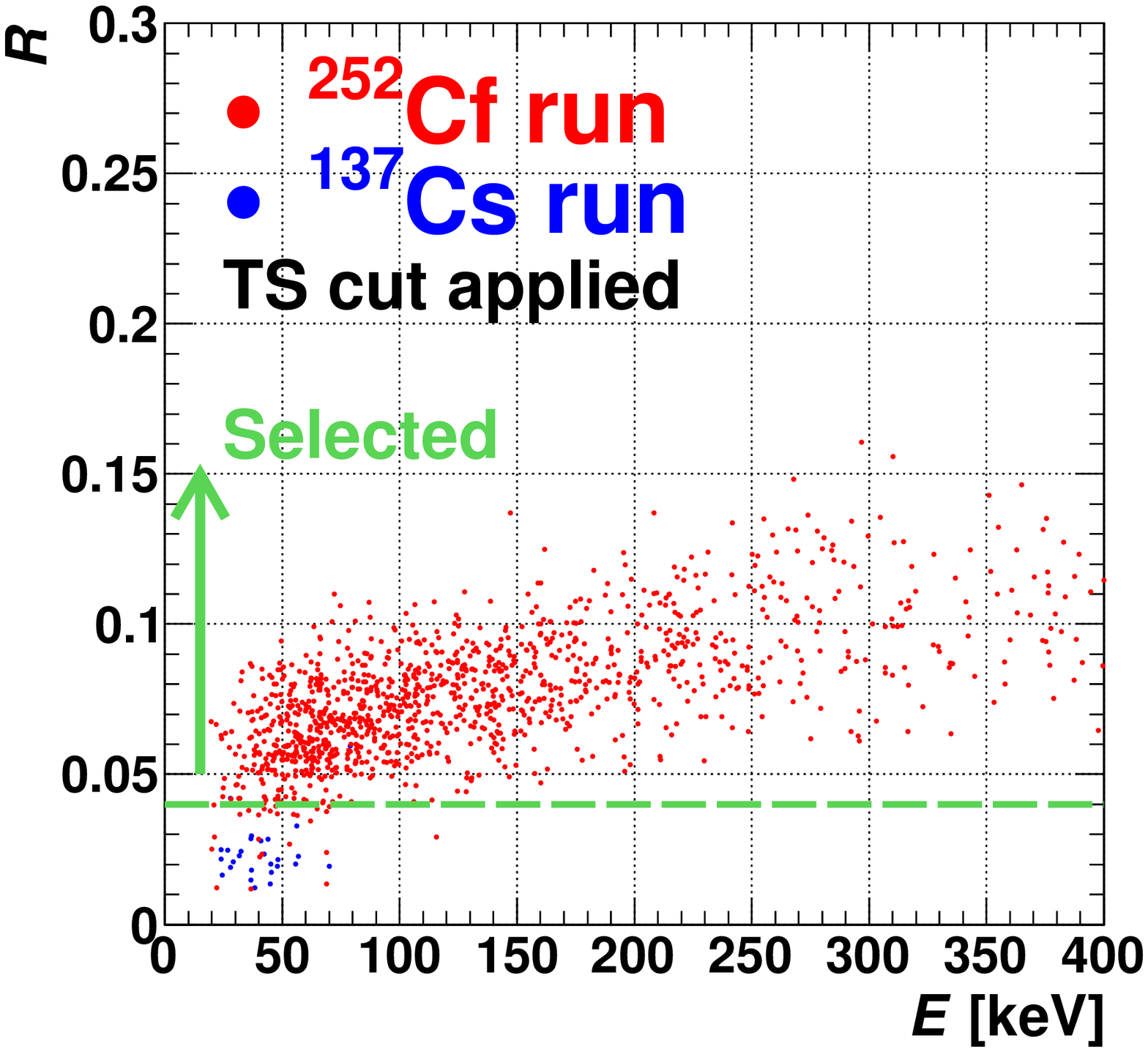}
        \hspace{1.6cm}
      \end{center}
    \end{minipage}
    
    \end{tabular}
    \caption{Energy dependences of $L$ after the fiducial cut (top-left), $TS_{\rm X}$ after the L cut(top-right), $TS_{\rm Y}$ after the L cut (bottom-left), and $R$ after the TS cut (bottom-right). Red and blue points represent the results of $^{252}\mathrm{Cf}$ and $^{137}\mathrm{Cs}$ measurements, respectively.}
    \label{fig:selec}
  \end{center}
\end{figure}

\begin{figure}[htbp]
  \begin{center}
  \begin{center}
  \includegraphics[clip, width=6.0cm]{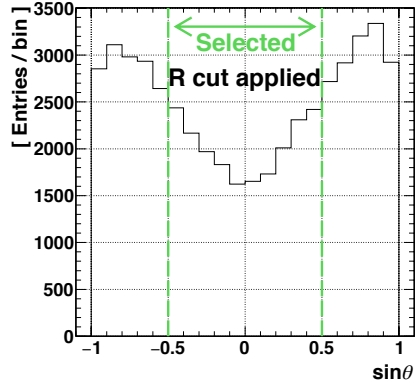}
  \hspace{1.6cm}
  \end{center}
  \caption{The measured sin$\theta$ distribution with $^{252}\mathrm{Cf}$ after the R cut.}
  \label{fig:sin_degree}
  \end{center}
\end{figure}

The energy dependencies of these parameters for nuclear recoil events by $^{252}\mathrm{Cf}$ and $\gamma$-ray events by $^{137}\mathrm{Cs}$ at each cut stage are shown in Fig.~\ref{fig:selec}.
(Top-left) Nuclear recoil tracks have shorter track length than electron tracks do because of a larger linear energy transfer.
Electron tracks in the137Cs run (blue points) are seen as a vertical band below $\sim$ 100~keV.
Nuclear recoil tracks are also expected in the $^{252}\mathrm{Cf}$ run.
Nuclear recoil tracks are seen as a band structure with shorter $L$s.
Therefore, the region below the green dashed line is selected. 
(Top-right and bottom-left)
Nuclear recoil tracks have larger $TS$s than electron tracks do because of a larger linear energy transfer.
Electron tracks in the $^{137}\mathrm{Cs}$ run (blue points) are seen in the low energy and low $TS_{\rm X}$($TS_{\rm Y}$) area.
Nuclear recoil tracks are also expected in the $^{252}\mathrm{Cf}$ run.
Nuclear recoil tracks are seen as a band structure with larger $TS$s.
Therefore, the regions above the green dashed lines are selected. 
(Bottom-right) 
Events close to the detection plane have smaller $R_{\rm X}$($R_{\rm Y}$) values than the ones with longer drift length because of the smaller gas diffusion.
Electron tracks in the $^{137}\mathrm{Cs}$ have already been rejected by the preceding cuts and here $\alpha$-ray background events from the $\mu$-PIC are seen.
Therefore, the regions below the green dashed lines are selected. 
The $\theta$ distribution for the $^{252}\mathrm{Cf}$ run after the R cut is shown in Fig.~\ref{fig:sin_degree}.
The $^{137}\mathrm{Cs}$ run and $^{252}\mathrm{Cf}$ run are runs with the corresponding radioactive sources.
A $^{137}\mathrm{Cs}$ source is used as a $\gamma$-ray source and a $^{252}\mathrm{Cf}$ source is used as a fast neutron and $\gamma$-ray source.

\begin{figure}[htbp]
  \begin{center}
  \begin{center}
  \includegraphics[width=150mm]{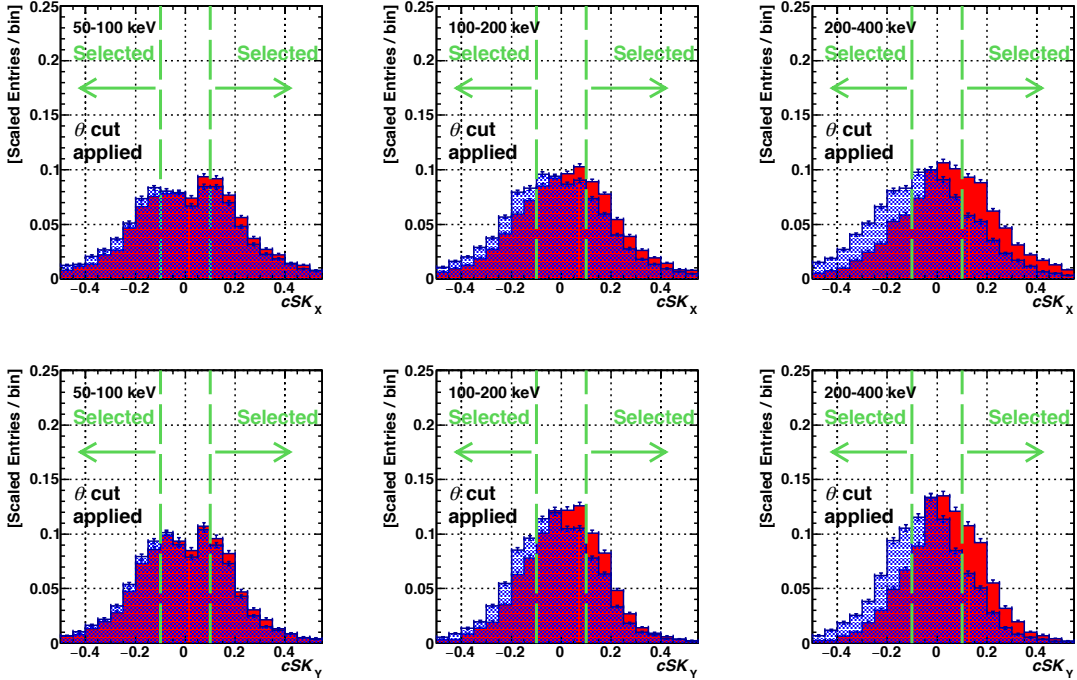}
  \hspace{1.6cm}
  \end{center}
  \caption{The measured $cSK_{\rm X}$ distributions for $\bf +X$ irradiation (blue) and $\bf -X$ irradiation (red) data in the upper panel and the $cSK_{\rm Y}$ distributions for $\bf +Y$  irradiation (blue) and $\bf -Y$ irradiation (red) data in the lower panel after the $\theta$ cut. Left, middle, and right panels correspond to 50-100, 100-200, and 200-400~keV, respectively.}
  \label{fig:skewness}
  \end{center}
\end{figure}

We then describe the 
detector performance specific to this analysis.
The first performance is the head-tail determination power ($HP$) or the sense determination of the measured tracks.
After the fundamental studies with prototype detectors \cite{9}, head-tail determination is applied to NEWAGE dark matter analysis for the first time.
The $HP$ is studied by irradiating the detector with neutrons from the neutron source at various positions.
Comparison of the $cSK$ distribution of irradiations with the source placed at the opposite positions provided the $HP$ results.
Because a head-tail determination along at least one axis can provide a head-tail to the track, $HP$s in the X$-$Y plane is studied.
A comparison of the results with 
the source at ($-25.5$, 0, 0) (hereafter referred to as $\bf -X$) and (25.5, 0, 0) ($\bf +X$) is used to evaluated the $HP$ along the X axis.
The data with (0, $-25.5$, 0) ($\bf -Y$) and (0, 25.5, 0) ($\bf +Y$) are used to evaluate the $HP$ along the Y axis.
The measured $cSK$ distributions are shown in Fig.~\ref{fig:skewness} for three energy ranges.
The upper panels show the $cSK_{\rm X}$ distributions for $\bf +X$ irradiation (blue) and $\bf -X$ irradiation (red) data and the lower panels show the $cSK_{\rm Y}$ distributions for $\bf +Y$ irradiation (blue) and $\bf -Y$ irradiation (red) data.
The $HP$ for an irradiation is defined as the fraction of normalized area
with the absolute values corresponding to $cSK$ larger than 0.1.
Binomial errors are assigned for the $HP$s.
The measured $HPs$ are summarized in Table \ref{tab:thtp}.
The results of $\bf +X$, $\bf -X$, $\bf +Y$ and $\bf -Y$ irradiation for each energy range are 
found to be consistent with each other within the statistical errors so the averaged values with propagated errors are going to be used in the following discussions.
The obtained $HP$s are (53.4$\pm$~0.5)\% for 50-100~keV, (57.7$\pm$0.4)\% for 100-200~keV and (65.1$\pm$0.5)\% for 200-400~keV, respectively.

\begin{table}[tbh]
\caption{Head-tail determination powers for 50-100 keV, 100-200 keV, 200-400 keV}
\begin{tabular}{|c|c|c|c|c|c|}
\hline
energy range & $HP_{\bf +X}$ [\%] & $HP_{\bf -X}$ \% & $HP_{\bf +Y}$ [\%] & $HP_{\bf -Y}$ [\%] & average [\%] \\ \hline \hline
50-100~keV & 52.2~$\pm$~0.8 & 55.4~$\pm$~0.9 & 54.0~$\pm$~0.9 & 52.0~$\pm$~1.2 & 53.4~$\pm$~0.5\\ \hline
100-200~keV & 57.7~$\pm$~0.7 & 57.4~$\pm$~0.8 & 59.3~$\pm$~0.8 & 56.5~$\pm$~1.1 & 57.7~$\pm$~0.4\\ \hline
200-400~keV & 65.2~$\pm$~0.8 & 63.9~$\pm$~0.9 & 67.5~$\pm$~0.8 & 63.6~$\pm$~1.2 & 65.1~$\pm$~0.5 \\ \hline \hline
histogram in Fig.~\ref{fig:skewness}& blue(top) & red(top) & blue(bottom) & red(bottom) & \\ \hline
\end{tabular}
\label{tab:thtp}
\end{table}

\begin{figure}[htbp]
  \begin{center}
    \begin{tabular}{c}

      \begin{minipage}{0.5\hsize}
        \begin{center}
          \includegraphics[clip, width=6.0cm]{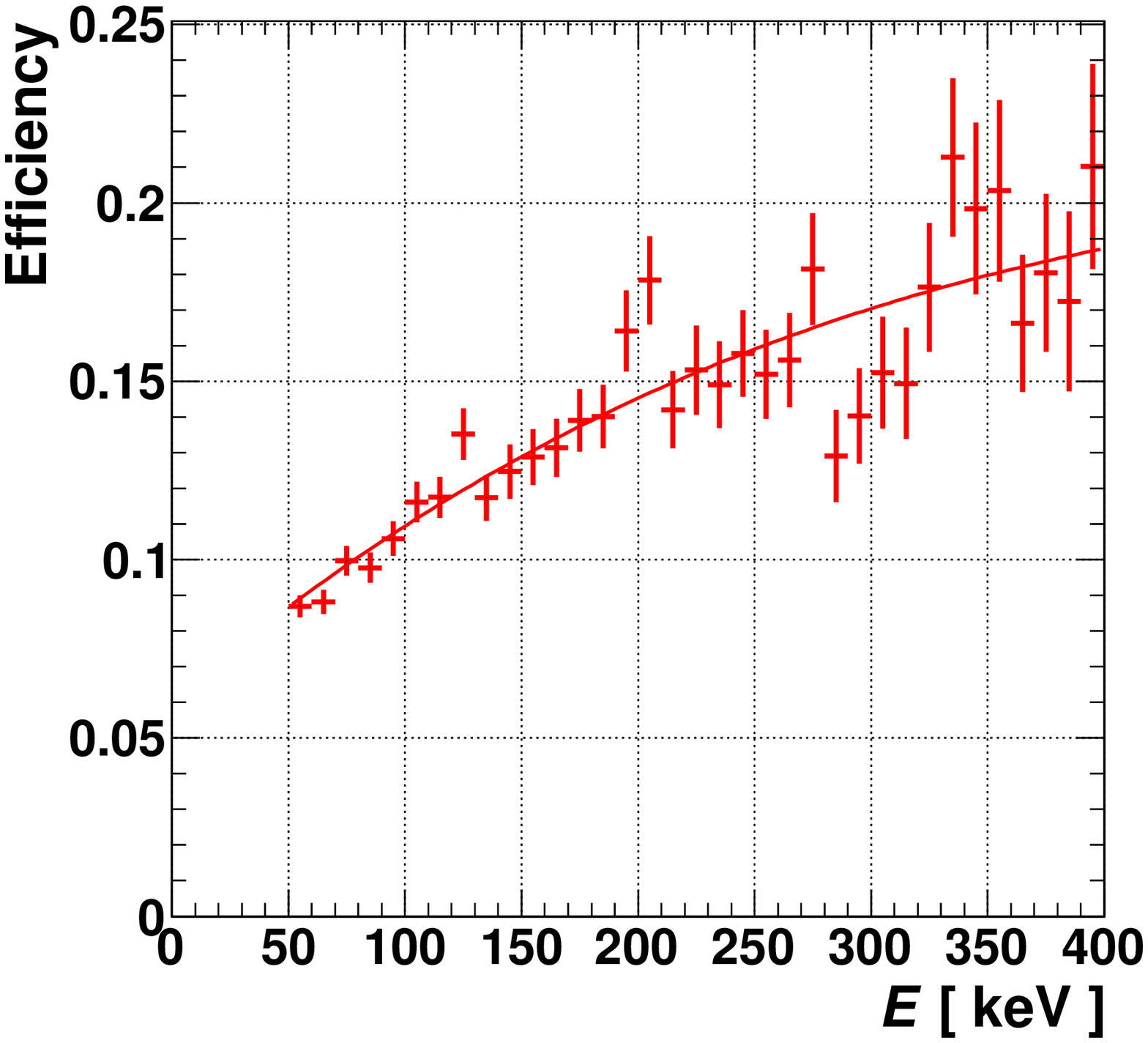}
          \hspace{1.6cm}
        \end{center}
      \end{minipage}

      \begin{minipage}{0.5\hsize}
        \begin{center}
          \includegraphics[clip, width=6.0cm]{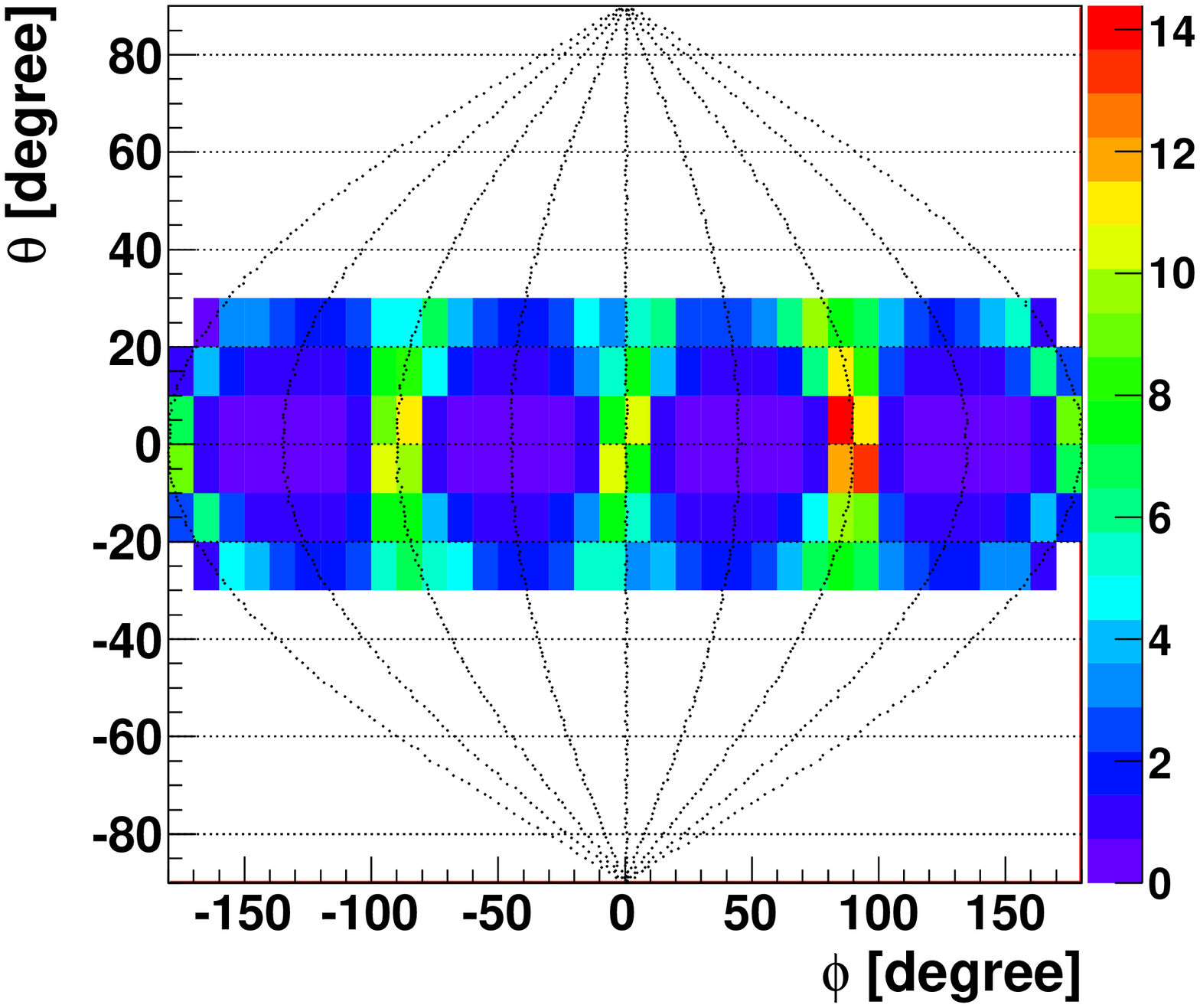}
          \hspace{1.6cm}
        \end{center}
      \end{minipage}

    \end{tabular}
    \caption{The detection efficiency(left) and directional response(right) of nuclear recoil events.}
    \label{fig:efficiency}
  \end{center}
\end{figure}

The detection efficiencies of nuclear and electron events after these cuts applied are 
evaluated by dividing the measured energy spectrum by the simulated one.
The denominator is the expected number of nuclear recoils simulated by the simulation.
An averaged spectrum of 6 measurements by placing a $^{252}\mathrm{Cf}$ to six positions is 
used to cancel the position dependence and to measure the overall response of the detector.
The six positions are $(25.5, 0 , 0)$, $(-25.5, 0, 0)$, $(0, 25.5, 0)$, $(0, -25.5, 0)$, $(0, 0, 47.5)$ and $(0, 0, -47.5)$.
Typical detection efficiencies for the nuclear recoil events after all cuts are shown in Fig.~\ref{fig:efficiency}(left).
The lower (50~keV) boundary is the energy threshold in our detector and the upper boundary (400~keV) is decided where the detector lose efficiency for even high mass ($>$ 1~TeV) WIMPs.
The detection efficiency of nuclear recoil events is found to be 9\% at 50~keV.
The detection efficiencies of electron events, or the gamma-ray rejection power, are evaluated by irradiating the detector with $\gamma$-rays from a $^{137}\mathrm{Cs}$ source and comparing the data with the simulation results.
The obtained detection efficiency of the electron events in the energy bin 50 - 100~keV is 1.25 $\times 10^{-5}$.

The directional response to an isotropic distribution of the track directions is shown in Fig.~\ref{fig:efficiency}(right).
This distribution is obtained as the sum of the distributions of six positions weight by livetimes and then normalized so that the mean equals 1.
The measured directional response is used for weighting the expected direction distribution of nuclear recoil tracks in the directional dark matter search analysis with a 3d-vector tracking analysis.

The angular resolution of the nuclear recoil tracks is measured using the fast neutrons from a ${}^{252}\mathrm{Cf}$ source.
It is evaluated by comparing the measured and simulated distributions of recoil angle.
$^{252}\mathrm{Cf}$ run data is used as the measured distributions, while simulated distribution is made with GEANT4 without the consideration of the angular resolution.
Simulated distribution is then smeared with various angular resolutions and the smeared distributions are compared with measured ones.
Angular resolution is determined by the best-fit distribution.
The measured angular resolution in the energy bin 50-100~keV is $(36\pm4)^{\circ}$\cite{15}.
The energy resolution is estimated as the convolution of the non-homogeneous detector response (gain and charge collection within the detection volume) term and the electronics-noise term.
The former is estimated from the width of the $^{220}$Rn and $^{222}$Rn peaks.
The latter is evaluated with the FADC waveform data of the off-trigger timing.
The radon decay events in the TPC gas make peaks around 6~MeV.
There are two isotopes of radon, $^{220}$Rn and $^{222}$Rn.
Energies of $\alpha$-rays from $^{220}$Rn and its progeny are 6.288~MeV, 6.779~MeV, 6.051 ~MeV (35.94\%) and 8.785 MeV (64.06\%), and those of $^{222}$Rn are 5.490~MeV, 6.003~MeV and 7.687~MeV.
The energy resolution is estimated by comparing the measured energy spectrum with the energy spectrum made by smearing these peaks from radons with various energy resolutions.
This peak by the radons in the energy spectrum is referred to as the radon peak afterwards.
The deterioration of the energy resolution is due to the position dependence of the gas gain and the attachment of electrons during the drift. 
The electric noise component in the energy resolution is evaluated from the fluctuation of the baseline of FADC waveform.
The FADC data for 100 clock in prior to the event signal is used and the standard deviation of the voltage is taken as the electric noise component in the energy resolution.
The obtained total energy resolution is 13 $\pm$ 1$\%$.
The energy resolution does not have energy dependence because the dominant cause of the energy resolution deterioration, position dependence of the gas gain and the attachment of electrons during the drift, does not have energy dependence. 

\clearpage


\section{Directional dark matter search}
\label{sec:dds}

\subsection{Measurement}
\label{sec:dataset}
A dark matter search experiment was carried out from July 2013 to August 2017 in Laboratory B, Kamioka Observatory
(36$^{\circ}$25' N, 137$^{\circ}$18' E), at a water equivalent depth of 2700~m.
The run properties are summarized in Table.~\ref{tab:t2}.
The ''main'' run number is incremented when the hardware is modified and the ''sub'' run number is incremented when the chamber gas is changed. The total livetime used for this work is 434.85~days, corresponding to an exposure of 4.51~kg$\cdot$days.

\begin{table}[tbh]
\caption{Measured dates and livetimes for this work. Detector orientation indicates the orientation of the Z axis of the detector from South or North to East. Gas flow rate is the rate of the gas circulation.}
\centering
\begin{tabular}{|l|l|l|l|l|}
\hline
Run number & Date & livetime& Detector& Gas circulation\\ 
(main-sub) & & [days]&orientation &rate[mL/min]\\

\hline \hline
Run14-1 & 2013/7/17   to 2013/9/16   & 17.10  & S60E &  500\\ \hline
Run14-2 & 2013/10/17 to 2013/11/14 & 14.52  & S60E &  500\\ \hline
\hhline{|=|=|=|=|=|}
Run14-3 & 2014/01/29 to 2014/3/12   & 25.34  & S60E &  500\\ \hline
Run16-1 & 2016/1/14   to 2016/3/10   & 42.28   & S60E &  1000\\ \hline
Run16-2 & 2016/3/25   to 2016/6/28   & 69.94  & S60E &  1000 \\ \hline
Run17-1 & 2016/6/28   to 2016/8/24   & 26.16  & S60E &  500\\ \hline
Run18-2 & 2016/9/1     to 2016/10/19 & 41.43  & N76E &  700  \\ \hline
Run18-3 & 2016/10/20 to 2017/1/19   & 66.86  & N76E &  700 \\ \hline
Run18-4 & 2017/1/26   to 2017/4/21   & 49.51  & N76E &  700\\ \hline
Run18-5 & 2017/4/27   to 2017/8/8     & 81.71 & N76E &  700\\ \hline
\hhline{|=|=|=|=|=|}
Total & 2013/7/17 to 2017/8/8 & 434.85  &&\\
\hline
\end{tabular}
\label{tab:t2}
\end{table}

Initial results from Run14-1 and Run14-2 data with 
a livetime of 31.62~days were previously reported \cite{6}.
Since then, additional data corresponding to 403.23~live-days have been accumulated.
The total exposure is about 14 times of that of the initial results.
The $Z$ axis of the detector was aligned to S30E for the first half and to S76E for the second half to minimize the potential systematic uncertainties.
The gas circulation rate was changed a few times as listed in the table aiming to check the effect on the gas gain stability and radon background which was found to be a dominant background source in our previous measurement\cite{16}.
No significant effect due to either factor is observed.   
The data accumulated in Runs~15 and 18-1 are not used for the analysis because the system suffered from electronic noise and the DAQ system was out of condition in the corresponding runs, respectively.


\clearpage
\subsection{Detector stability and data correction}
\label{detector_sta}
The detector gas was filled at the beginning of each sub-run and then it was circulated without any change during the sub-run. 
Because the detector performance changes mainly due to the deterioration of the chamber gas, the detector performance was monitored for the data correction.
This stability study and data correction are newly introduced for this directional dark matter search since the typical data-taking period without gas change is longer than that in previous runs.


\subsubsection{Gas gain correction}
\label{sec:gain}
\begin{figure}[htbp]
   \vspace*{-0.2cm}
    \begin{center}
       \includegraphics[height=80mm,width=110mm]{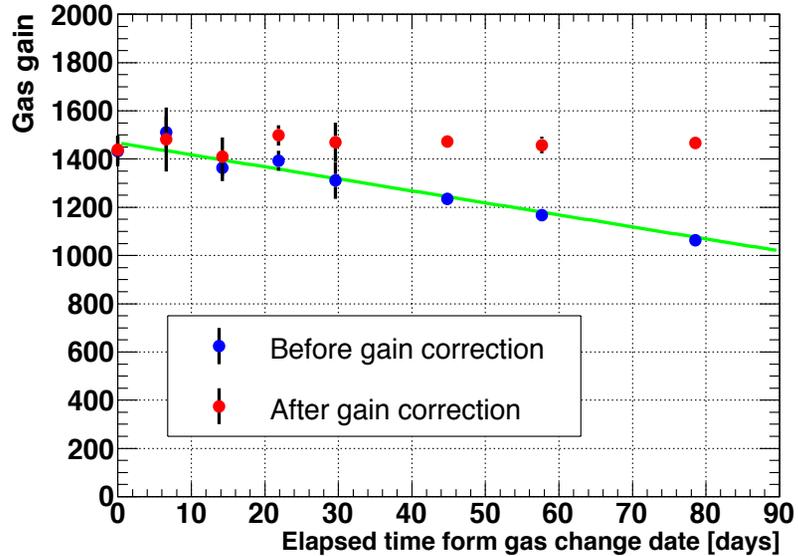}
    \end{center}
    \caption{Time dependence of gas gains before (blue) and after (red) correction.}
    \label{fig:Corrected_gain}
\end{figure}

The gas gain was monitored throughout the measurement period by the amount of detected charge of $\alpha$-rays from the radon progeny.
The detected charge is the product of the primary charge and the gas gain, where the primary charge is known.
Observed gas gains as a function of the elapsed time are shown in Fig.~\ref{fig:Corrected_gain} with blue points. 
The gas gain is observed to decrease $\sim 10~\%$/month.
The decrease is due to the out-gas and/or the leaks of the chamber.
The gains are fitted with a linear function and the conversion factors are corrected so that the obtained line became constant. 
The corrected gas gains are also shown in Fig.~\ref{fig:Corrected_gain} with red points. The event energy is corrected before the event selection.

\clearpage
\subsubsection{TS correction}
\label{sec:TOTsum}
\begin{figure}[htbp]
    \begin{center}
       \includegraphics[height=70mm,width=150mm]{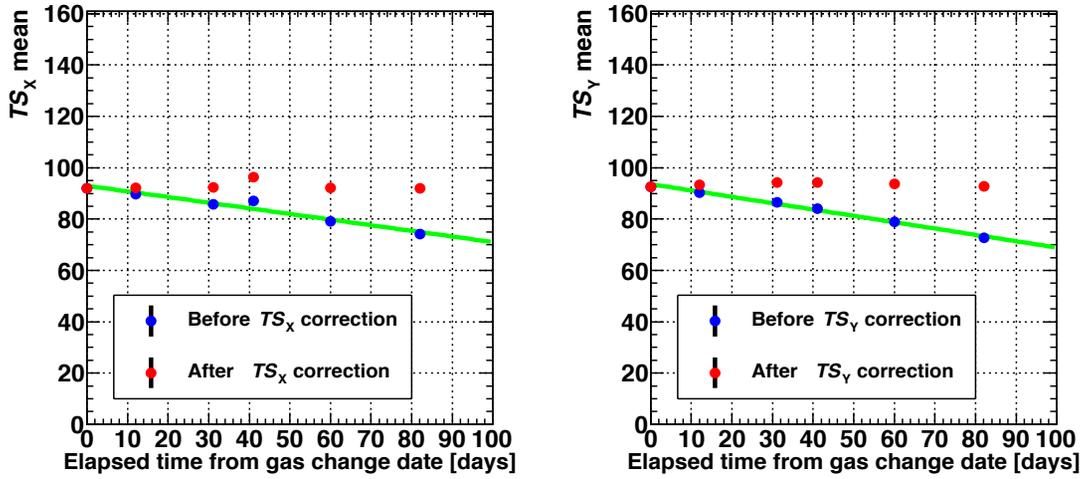}
    \end{center}
    \caption{The time transition of the $TS$s for the events in the 50-60~keV energy bin. The $TS$ of X and Y are shown in the left and right panels, respectively. The blue and red marks indicate the $TS$ before and after corrections, respectively.}
    \label{fig:TOTsum_Before}
\end{figure}

$TS$s are also observed to decrease as a function of time, mainly due to the gas gain decrease.
The signal at each strip becomes smaller with a decreased gas gain and thus the $TS$s decrease.
This is an independent observable from the energy information.
The energy is known from the ``charge" and the $TS$s are known from the ``track".
$TS$s need to be corrected in a different way from the energy for the gas gain decrease.
Time dependence of the $TS$ is studied and used for the correction to recover the inefficiency due to the decrease of the $TS$.
The time dependence of the mean values of $TS_{\rm X}$  and $TS_{\rm Y}$ for the 50-60~keV energy-bin are shown in Fig.~\ref{fig:TOTsum_Before} with blue points.
The $TS$s are corrected in the same manner as for the gain correction and the corrected ones are shown with red points. The correction functions are prepared for each 10 keV energy bin and the $TS$s are corrected before the TS cuts.

\clearpage
\subsection{Results}
\label{sec:dmsearch}
\begin{figure}[htbp]
    \begin{center}
       \includegraphics[width=100mm]{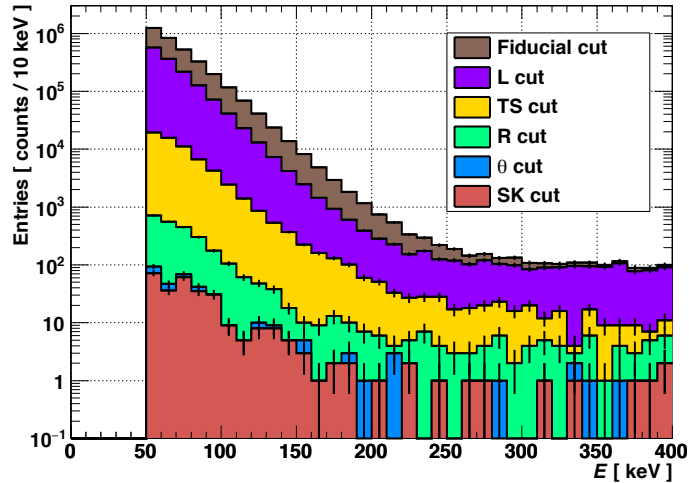}
    \end{center}
    \vspace*{-0.5cm}
    \caption{Obtained energy spectra in this work at each cut-step. The unit of the vertical axis is counts/10 keV.} 
    \label{fig:ene_cts}
\end{figure}

\begin{figure}[htbp]
    \begin{center}
    \includegraphics[width=100mm]{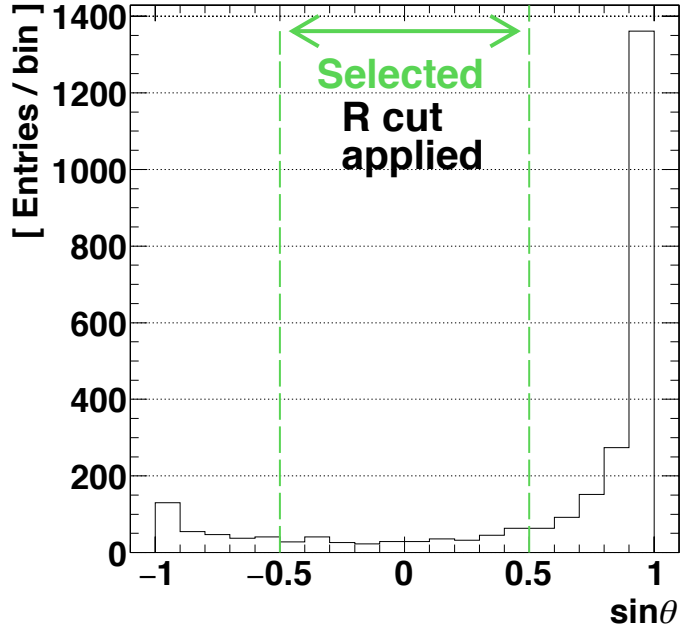}
    \end{center}
    \vspace*{-0.5cm}
    \caption{The $\sin \theta$ distribution of measured tracks in the energy range of 50 - 400~keV after the R cut.}
    \label{fig:dm_skymap_before}
\end{figure}

Event selections described in Sec.~\ref{sec:evtsel} are applied to the collected data.
Energy spectra at each selection step are shown in Fig.~\ref{fig:ene_cts}.
The lower energy bound is set to 50~keV, below which no direction sensitivity is measured.
The upper energy bound is set to 400~keV with consideration of the recoil energy spectrum by WIMPs of 1~$\rm TeV/c^2$.
It can be seen that the L cut rejected $\gamma$-rays below 200~keV as explained in Sec.\ref{sec:evtsel}.
It can be seen that the L cut rejected $\gamma$-rays below 200~keV.
The TS, R, and $\theta$ cuts are effective throughout the energy range of interest.
A $\sin \theta$ distribution before the $\theta$ cut is shown in Fig.~\ref{fig:dm_skymap_before} (corresponding to the green spectrum in Fig.~\ref{fig:ene_cts}) to demonstrate the effect of the newly introduced $\theta$ cut.
Our previous study indicated that the $\mu$-PIC is contaminated with radioactive isotopes such as 
$\rm^{238}U$ and $\rm^{232}Th$.
Alpha-rays are emitted from these radioactive isotopes, causing background events.
These up-going background events are found at a peak around one in the $\sin\theta$ distribution.
They are effectively rejected by the $\theta$ cut.
A total reduction of four orders of magnitude is realized at 50~keV, whereas the detection efficiency of the nuclear recoil is retained at $\sim10\%$ as discussed in Sec.~\ref{sec:evtsel}.

\begin{figure}[htbp]
    \begin{center}
       \includegraphics[width=100mm]{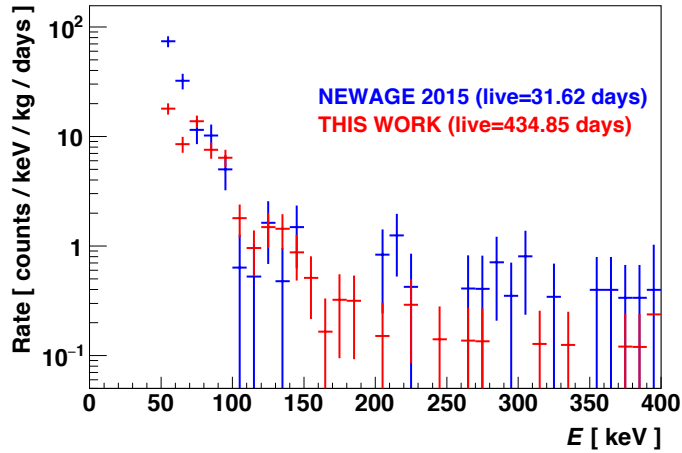}
    \end{center}
    \vspace*{-0.5cm}
    \caption{Measured ''effective'' energy spectrum (red histogram). The result of  the previous work is shown with blue histogram\cite{6}. Both spectra are unfolded with the detection efficiency of nuclear recoil tracks for comparison.}
    \label{fig:spec}
\end{figure}

\begin{figure}[htbp]
    \begin{center}
       \includegraphics[width=100mm]{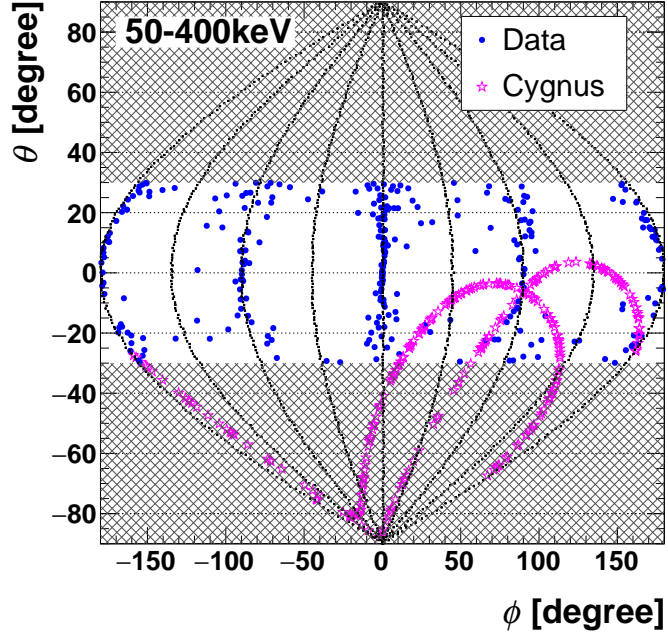}
    \end{center}
    \vspace*{-0.5cm}
    \caption{The directional distribution of measured tracks in the energy range of 50 - 400~keV.}
    \label{fig:dm_skymap}
\end{figure}

The ''effective'' energy spectra of this work and our previous work are shown in Fig.~\ref{fig:spec}. The ''raw'' energy spectra after all cuts are unfolded with the nuclear detection efficiencies so that the effective energy spectra could be compared to one another.
It is seen that the effective count rate is reduced by a factor of 4 from the previous run at 50~keV. This is due to the newly introduced $\theta $ cut.
The sky-map after all cuts is shown in Fig.~\ref{fig:dm_skymap}.
The direction of each event is plotted with a blue point with the corresponding direction of the constellation Cygnus. Two trajectory loops for the Cygnus direction are seen indicating the orientation change of the detector between Runs~17 and Runs~18. 
The gray hatched areas are the cut area and some Cygnus directions corresponding to the events in the selected area are shown there.

\subsection{Systematic Errors}
\label{sec:systematicerror}
The systematic errors relevant to this analysis are summarized in Table~\ref{tab:tsyserr}. Because the directional analysis is performed by comparing the measured and expected distributions of the angle between the recoil direction and direction of the WIMP-wind, or the $\cos \theta_{\rm Cyg}$ spectra, the effect of the systematic errors on the expected cos$\theta_{\rm Cyg}$ spectrum are studied.
Differences of the expected number of event with and without the consideration of the systematic errors divided by the expected number of events without the consideration of the systematic error are listed as the effect on shape of cos$\theta_{\rm Cyg}$ spectrum in Table~\ref{tab:tsyserr}.  
The energy scale and angular resolution would mainly change the shape of the spectrum. The head-tail determination also affects the shape, whereas its effect is found to be small compared with the angular resolution. 
The energy resolution would change the total rate of the $\cos\theta_{\rm Cyg}$ spectrum.
These errors will be used in the following analysis.

\begin{table}[tbh]
\caption{Systematic errors for angular resolution, energy resolution and head-tail determination and energy scale. The effects on the shape of cos$\theta_{\rm Cyg}$ spectrum is defined as the relative change of number of events in one bin.}
\centering
\begin{tabular}{|c|c|c|c|}
\hline
Systematic error & mean & error & effect on shape of \\
 & & (1~$\sigma$) & cos$\theta_{\rm Cyg}$ spectrum \\ \hline \hline
Angular resolution & $36^\circ$ & $4^\circ$ & 7.4\% \\  \hline
Energy resolution & 13\% & 1\% & 1.5\% \\  \hline
Head-tail determination & 53.4\% & 0.5\% & 1.4\% \\  \hline
Energy scale & 0\% & 5\% & 11.4\% \\  \hline
\end{tabular}
\label{tab:tsyserr}
\end{table}

\clearpage

\subsection{Dark matter limits}
\label{dmlimit}
Limits on WIMP-nucleon cross section are obtained by a 3d-vector directional analysis. Here 3d-vector directional analysis means a directional dark matter search analysis using 3d nuclear recoil track information with and head-tail sensitivities.
Astrophysical parameters, nuclear parameters, and detector responses are listed in Table~\ref{tab:tpara}.
The main scheme of this method serves to compare 
measured and expected $\cos \theta_{\rm Cyg}$ spectra. 
Measured $\cos \theta_{\rm Cyg}$ with raw number of events are compared with the expected $\cos \theta_{\rm Cyg}$ spectrum considering the detector response.

Most of the methods used to get expected $\cos \theta_{\rm Cyg}$ distribution are unchanged from our previous work~\cite{6}.
The main differences are that four bins in $\cos \theta_{\rm Cyg}$ covering $-1$ to 1 (previous two bins from 0 to 1) are used and
a binned likelihood-ratio method is adopted\cite{17,18}.

\begin{table}[tbh]
\caption{Astrophysical parameters, nuclear parameters, and detector responses used for the 3d-vector directional analysis}
\centering
\begin{tabular}{|l|l|}
\hline
WIMP velocity distribution & Maxwellian  \\ \hline
Maxwellian velocity dispersion  &  $v_0$ = 220 km/s\\  \hline
Escape velocity &  $v_{esc}$ = 650 km/s\\  \hline
Local halo density& $\rho_{\rm DM}$ = 0.3 GeV/$c^2/{\rm cm^3}$  \\  \hline
 Spin factor of ${}^{19}$F & $\lambda^2J(J + 1)$ = 0.647 \\  \hline
 Energy resolution at 50 keV& $(6.5\pm0.5)$ keV \\  \hline
Angular resolution at 50 - 100 keV & $(36\pm4)^\circ$ \\  \hline
Head-tail Precision at 50 - 100 keV & $(53.4\pm0.5)\%$\\ \hline
Energy scale & $(0\pm5)\%$\\ \hline
\end{tabular}
\label{tab:tpara}
\end{table}

A $\chi^2$ value for a given WIMP mass and energy bin is defined as

\begin{eqnarray}
\label{equ:chi2}
  \chi^2(\sigma^{\rm SD}_{\chi-{\rm p}}) & = & 2\sum^3_{i=0}\left[\{ N^{\rm exp}_i(\sigma^{\rm SD}_{\chi-{\rm p}})-N^{\rm data}_i\}+N^{\rm data}_{i}{\rm ln}\frac{N^{\rm data}_{i}}{N^{\rm exp}_i(\sigma^{\rm SD}_{\chi-{\rm p}})}\right]+\sum^3_{j=0}\alpha^2_{j}\\
  \alpha_{j}&=&\frac{\xi_{j}}{\sigma_{j}}
\end{eqnarray}

Here, the subscript $\it i$ is the bin number of the $\cos\theta_{\rm Cyg}$ distribution and the subscript $\it j$ is the type of the systematic errors ($\it j$=0, 1, 2 and 3 correspond to the angular resolution, the energy resolution, the head-tail determination and energy scale, respectively).
$N^{\rm exp}_i(\sigma^{\rm SD}_{\chi-{\rm p}})$ is the expected number of events for the WIMP-proton SD cross section of $\sigma^{\rm SD}_{\chi-{\rm p}}$, and $N^{\rm data}_i$ is the number of observed events.
Nuisance parameters $\alpha_{j}$ are introduced to consider the systematic errors.
Here $\xi_{j}$ and $\sigma_{j}$ are the shift from the central value and the systematic errors listed in Table~\ref{tab:tsyserr}, respectively.

\begin{figure}[htbp]
  \begin{center}
    \begin{tabular}{c}

      \begin{minipage}{0.5\hsize}
        \begin{center}
          \includegraphics[clip, width=7.0cm]{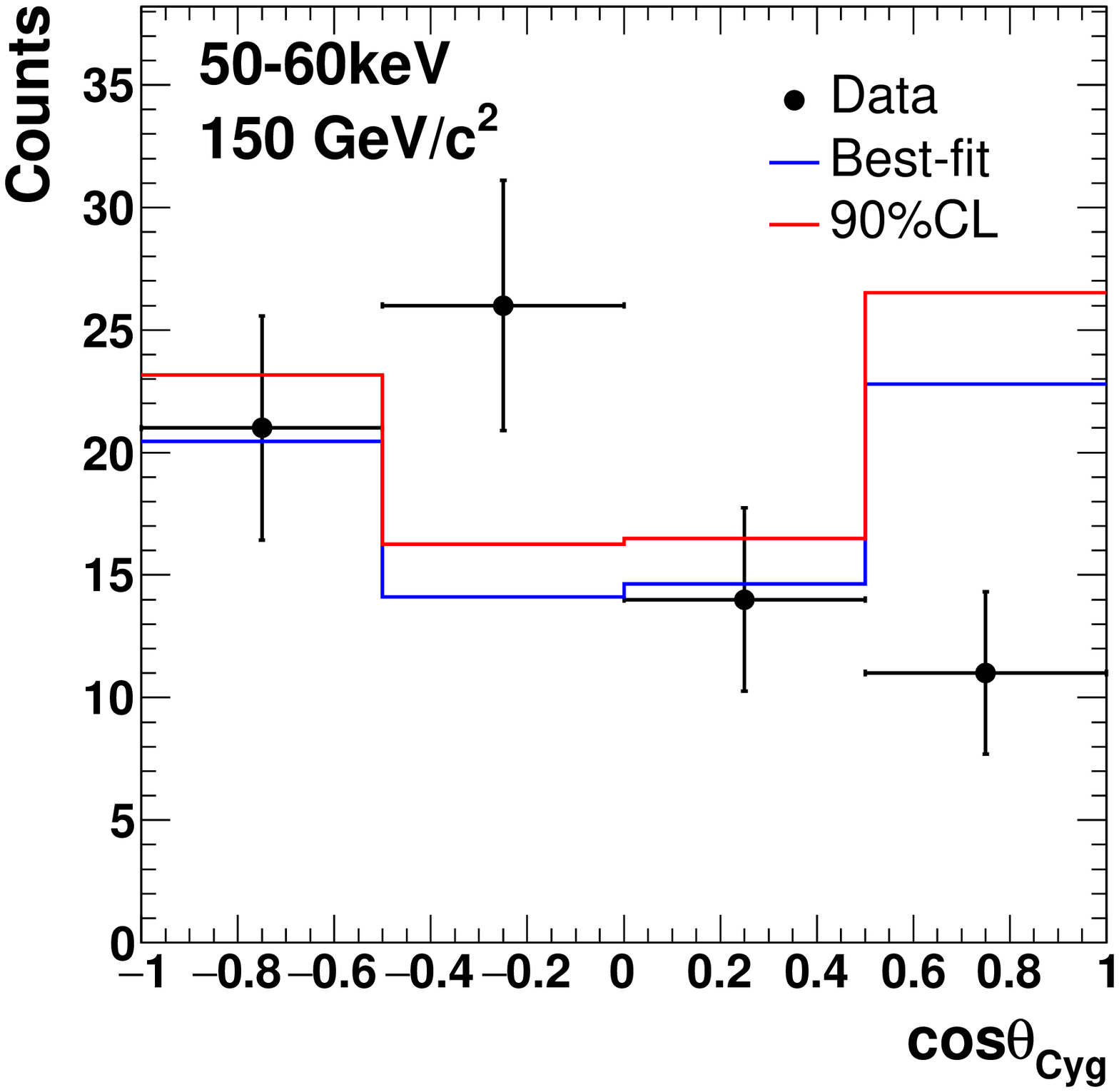}
          \hspace{1.6cm}
        \end{center}
      \end{minipage}

      \begin{minipage}{0.5\hsize}
        \begin{center}
          \includegraphics[clip, width=7.0cm]{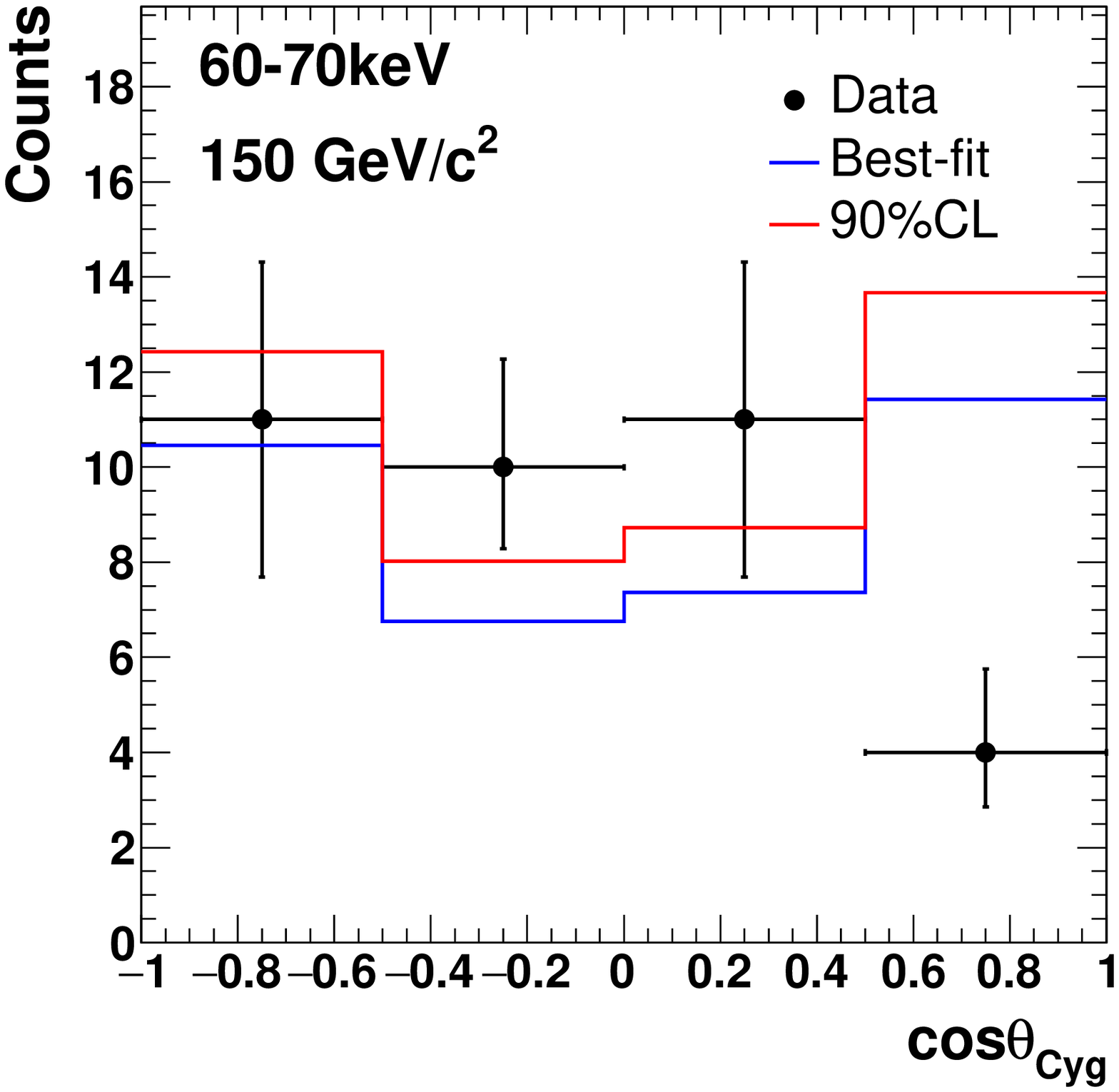}
          \hspace{1.6cm}
        \end{center}
      \end{minipage}

    \end{tabular}
    \caption{Measured (black) and calculated $\cos\theta_{\rm Cyg}$ distributions. The blue and red histograms are the best-fit and the 90\% C.L. ones, respectively. The energy bins are 50-60~keV and 60-70~keV for WIMP mass 150$\rm GeV/c^2$, respectively.}
    \label{fig:costheta_cyg_anal}
  \end{center}
\end{figure}

\begin{figure}[htbp]
  \begin{center}
    \begin{tabular}{c}

      \begin{minipage}{0.5\hsize}
        \begin{center}
          \includegraphics[width=8.0cm]{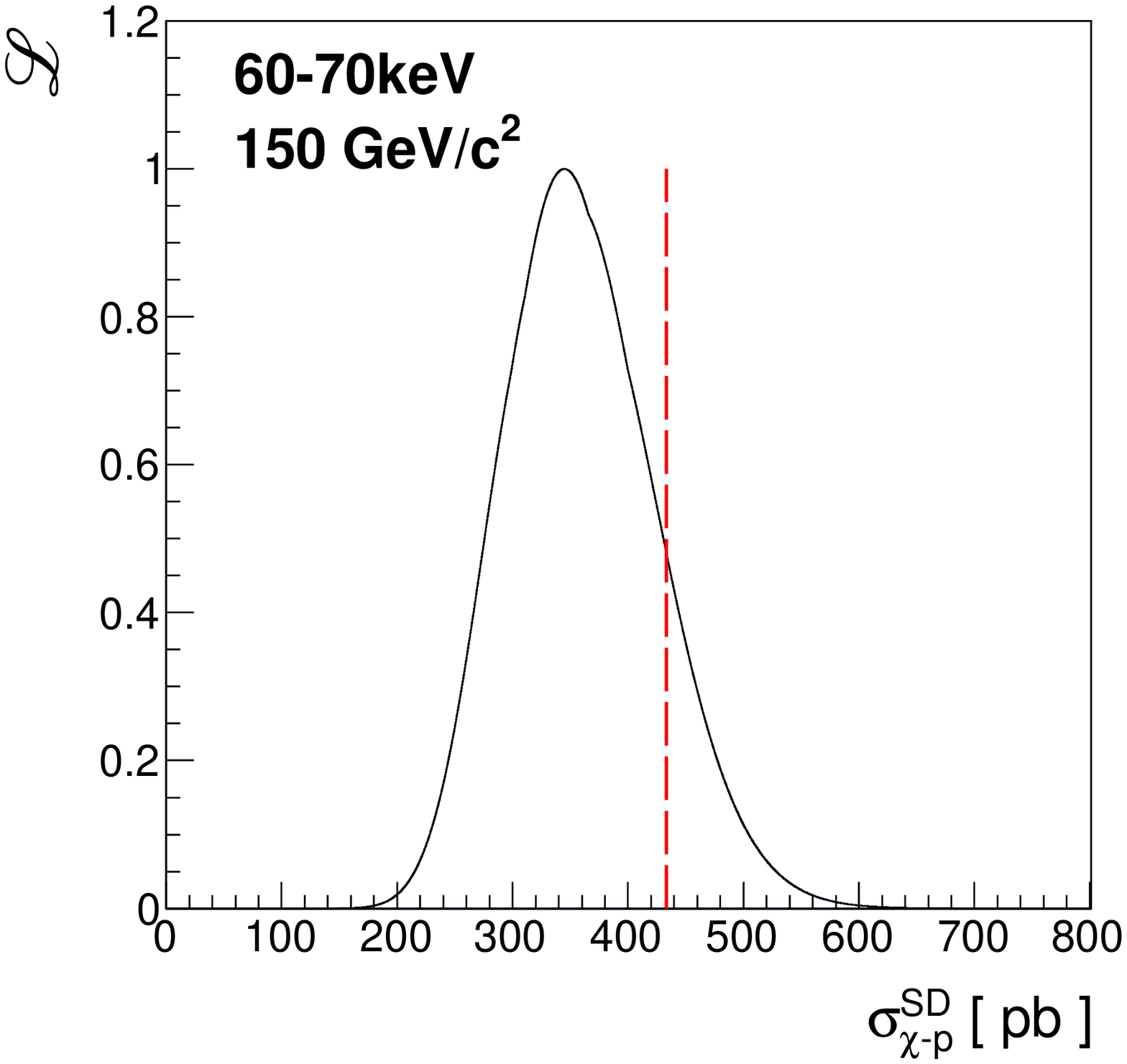}
          \hspace{1.6cm}
        \end{center}
      \end{minipage}

    \end{tabular}
    \caption{$\chi^2$ distribution at minimum $\alpha$ in the energy range of 60-70~keV for 150~GeV/$c^2$ WIMP. The 90\% C.L. upper limit of the $\sigma^{\rm SD}_{\chi-{\rm p}}$ is indicated by a red line.}
    \label{fig:L90}
  \end{center}
\end{figure}

The measured and best-fit (minimum $\chi^2$) $\cos\theta_{\rm Cyg}$ histograms for 50-60~keV and 60-70~keV energy ranges are shown in Fig.~\ref{fig:costheta_cyg_anal}.
The value of minimum $\chi^2$ over degree of freedom is 9.8/3 at $\alpha_0 = 0.6, \alpha_1 = 0$, $\alpha_2 = -0.1$, and $\alpha_3 = 0$.
Because no significant WIMP excess is found, 90\% confidence level (C.L.) upper limits are set on the WIMP-proton cross section.
A likelihood ratio $\mathscr{L}$ is defined by Eq. (\ref{equ:L}),
\begin{eqnarray}
\label{equ:L}
  \mathscr{L} = \rm exp\left(-\frac{\chi^2(\sigma^{\rm SD}_{\chi-{\rm p}}) - \chi^2_{min}}{2}\right),
\end{eqnarray}
where $\chi^2_{\rm min}$ is the minimum $\chi^2$ value. 
$\mathscr{L}$ values are shown in Fig.~\ref{fig:L90} as a function of $\sigma^{\rm SD}_{\chi-{\rm p}}$.

The 90\% C.L. upper limit is obtained using the relation defined by Eq.(\ref{equ:90CL}).

\begin{eqnarray}
\label{equ:90CL}
  \frac{\int_0^{\sigma^{\rm SD,limit}_{\chi-{\rm p}}} \mathscr{L} d\sigma^{\rm SD}_{\chi-{\rm p}}}{\int_0^{\infty} \mathscr{L} d\sigma^{\rm SD}_{\chi-{\rm p}}} = 0.9,
\end{eqnarray}
where $\sigma^{\rm SD,limit}_{\chi-{\rm p}}$ is the 90\% C.L. upper limit of the  $\sigma^{\rm SD}_{\chi-{\rm p}}$.
$\sigma^{\rm SD,limit}_{\chi-{\rm p}}$ is indicated with a red line in Fig.~\ref{fig:L90}.
An upper limit of $4.3\times10^{2}$~pb is obtained in this case.

\begin{figure}[htbp]
    \begin{center}
       \includegraphics[width=100mm]{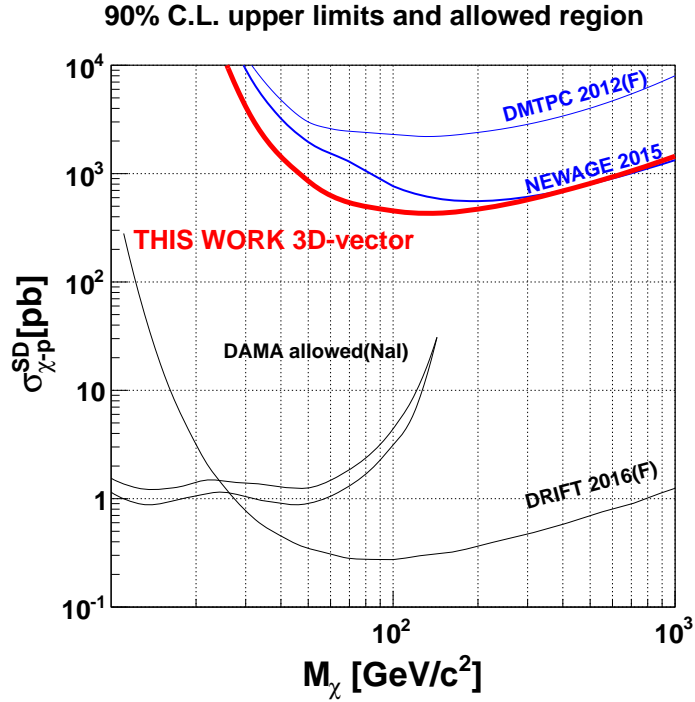}
    \end{center}
    \vspace*{-0.5cm}
    \caption{ 90$\%$ C.L. upper limits on the SD WIMP-proton cross section  as a function of WIMP mass. The red thick solid line is the result of this work (3d-vector directional analysis). The blue thin line labeled ``NEWAGE2015'' is our previous result. One of the interpretation of the DAMA's results is shown as ''DAMA allowed (NaI)''\cite{19}. Blue and black solid-lines show the  limits set by gas detectors with directional and conventional analysis, respectively\cite{20}.}
    \label{fig:limit}
\end{figure}
The 90$\%$ C.L. upper limit on SD WIMP-proton cross section 
obtained by scanning the WIMP mass and energy bin are shown with a red line in Fig.~\ref{fig:limit}. 
Limits with a 3d-vector tracking analysis is obtained for the first time by this work. 
The 90~\% C.L. upper limit on SD WIMP-proton cross-section of $4.3\times10^{2}$~pb for a 150~GeV/$c^2$ WIMP is obtained.
The directional limits below 150~GeV/$c^2$ WIMP mass are improved by this work owing to the newly-introduced $\theta$-cut.
Limits above 150~GeV/$c^2$ are similar to our previous limits due to the statistical fluctuation.

\section{Discussion}
\label{sec:discussion}
Three-dimensional trackings with head-tail sensitivities (3d-vector tracking analysis) are discussed as an ''ideal case'' for the directional dark matter search because it requires the smallest number of events to discover standard halo WIMPs together with its possibility for unexpected discoveries\cite{4,21,22,23}. 
This work demonstrated the first dark matter search with a 3d-vector tracking analysis.
In this study we use the skewnesses of the $TOT$ distributions of $X$ and $Y$ for the head-tail determination.
There is another axis, $Z$, detected as the time-evolution of the charge arrival (FADC wveform), which can also be used to determine the head-tails.
The waveform can be analyzed with $X$ and $Y$ parameters and this redundancy can be used to improve the head-tail determination power in the future analysis.
On the other hand as is seen in Fig.~\ref{fig:ene_cts}, background events limited the sensitivities of the 3d-vector directional dark matter search.
Therefore, it is necessary to reduce the background in addition to develop large-sized detectors.
The main background is found to be the $\alpha$-rays from the $\mu$-PIC \cite{6}.
Although the newly-introduced  $\sin\theta$ cut worked well, it is necessary to reduce the background itself at a hardware level.
A $\mu$-PIC with the less amount of contamination of radioactive isotopes by more than two orders of magnitude was developed and installed in NEWAGE-0.3b'\cite{24}.
Another interesting R\&D item which would accelerate the sensitivity improvement of the directional dark matter searches is the use of the negative-ion gas.
This type of gas, in which negative ions are drifted instead of electrons, first drew attention because of the small diffusion\cite{25}. 
Then it is demonstrated that some variations of this type of gases make the fiducialization in the Z-direction (drift-direction) possible.
This breakthrough is demonstrated first with $\rm CS_2$-based gas mixture and then pure $\rm SF_6$\cite{26,27} gas.
Intensive studies are being performed to implement this negative-ion TPC technology to the NEWAGE detector with a goal of background reduction of more than two orders of magnitude\cite{28}.
Large-volume negative-ion TPCs made of low-background materials would make the dark matter search possible even beyond the  neutrino floor where large-mass detectors without direction-sensitivity would rapidly lose their searching powers\cite{4}.

\section{Conclusions}
\label{sec:conclusion}
The first 3d-vector directional dark matter search   
using the NEWAGE-0.3b' detector
was performed. 
The search was carried out from July 2013 to August 2017 (Runs14 to Runs18).
The total livetime is 434.85~days corresponding to an exposure of 4.51~kg$\cdot$days which is about 14 times larger than that of our previous measurement (NEWAGE 2015).
A 90~\% C.L. upper limit on SD WIMP-proton cross-section of $4.3\times10^{2}$~pb for a 150~GeV/$c^2$ WIMP is obtained.
This is the first experimental dark matter limit obtained with a 3d-vector tracking analysis.


\section*{Acknowledgment}
We gratefully acknowledge the cooperation of Kamioka Mining and Smelting Company.
This work is supported by the Japanese Ministry of Education, Culture, Sports, Science and Technology, a Grant-in-Aid for Scientific Research, ICRR Joint-Usage, Japan Society for the Promotion of Science (JSPS) KAKENHI Grant Numbers 16H02189, 26104004, 26104005, 26104009, 19H05806 and the JSPS Bilateral Collaborations (Joint Research Projects and Seminars) program and Program for Advancing Strategic International Networks to Accelerate the Circulation of Talented Researches, JSPS, Japan(R2607).



\end{document}